\documentclass[preview,authoryear,3p,onecolumn]{elsarticle}
\usepackage{float,graphicx}
\usepackage{amssymb,amsmath,amsfonts,amstext,bm}
\usepackage{ifpdf,lineno,natbib,url}
\usepackage[colorlinks=true,breaklinks=true]{hyperref}
\usepackage{doi}
\usepackage{epstopdf}
\usepackage{graphics}
\usepackage{color}
\usepackage{psfrag}
\usepackage{calc}
\usepackage[english]{algorithm2e}
\usepackage{textcomp}
\usepackage{epstopdf}
\usepackage{accents}
\graphicspath{{./figures/}}
\numberwithin{equation}{section}
\usepackage{lineno}

\usepackage{empheq} 
\usepackage{mathtools} 
\usepackage{color,soul} %
\usepackage{bibentry} 
\usepackage{enumitem}
\usepackage{stmaryrd} 
\usepackage{multirow} 
\usepackage[font=normalsize]{subfig}

\usepackage{etoolbox}
\makeatletter
\def\@mkboth#1#2{}
\newlength\appendixwidth
\preto\appendix{\addtocontents{toc}{\protect\patchl@section}}
\newcommand{\patchl@section}{%
\settowidth{\appendixwidth}{\textbf{Appendix }}%
\addtolength{\appendixwidth}{1.5em}%
\patchcmd{\l@section}{1.5em}{\appendixwidth}{}{\ddt}%
}
\makeatother

\newcommand{\boldface}[1]{\boldsymbol{#1}}  

\newcommand{\bfd}{\boldface{d}}
\newcommand{\bfe}{\boldface{e}}

\newcommand{\bfn}{\boldface{n}}

\newcommand{\bfp}{\boldface{p}}

\newcommand{\bfu}{\boldface{u}}
\newcommand{\bfv}{\boldface{v}}
\newcommand{\bfw}{\boldface{w}}
\newcommand{\bfx}{\boldface{x}}

\newcommand{\bfC}{\boldface{C}}

\newcommand{\bfI}{\boldface{I}}

\newcommand{\bfR}{\boldface{R}}
\newcommand{\bfS}{\boldface{S}}

\newcommand{\bfeps}{\boldsymbol{\varepsilon}}
\newcommand{\bfepsvar}{\boldsymbol{\epsilon}}

\newcommand{\bfsigma}{\boldsymbol{\sigma}}

\newcommand{\bfvarphi}{\boldsymbol{\varphi}}

\newcommand{\bfnull}{\boldsymbol{0}}
\newcommand{\calB}{\mathcal{B}}

\newcommand{\calD}{\mathcal{D}}

\newcommand{\calI}{\mathcal{I}}

\newcommand{\calO}{\mathcal{O}}

\newcommand{\calS}{\mathcal{S}}
\newcommand{\calT}{\mathcal{T}}

\newcommand{\dsC}{\mathbb{C}}

\newcommand{\dsE}{\mathbb{E}}

\newcommand{\dsR}{\mathbb{R}}

\newcommand{\dsT}{\mathbb{T}}
\newcommand{\dsU}{\mathbb{U}}

\newcommand{\partderiv}[2]{\frac{\partial #1}{\partial #2}}

\newcommand{\T}{^{\mathrm{T}}} 

\newlength{\boxwidth}
\def\dd{\;\!\mathrm{d}}

\def\btheorem{\begin{theorem}}
\def\etheorem{\end{theorem}}
\def\blemma{\begin{lemma}}
\def\elemma{\end{lemma}}
\def\bproposition{\begin{proposition}}
\def\eproposition{\end{proposition}}
\def\bcorollary{\begin{corollary}}
\def\ecorollary{\end{corollary}}
\def\bdefinition{\begin{definition}}
\def\edefinition{\end{definition}}
\def\bexample{\begin{example}}
\def\eexample{\end{example}}
\def\bremark{\begin{remark}}
\def\eremark{\end{remark}}

\def\ol{\overline}

\DeclareMathOperator{\divv}{div}

\DeclareMathOperator{\curl}{curl}

\newcommand{\be}{\begin{equation}}
\newcommand{\ee}{\end{equation}}

\newcommand{\beq}{\begin{eqnarray}}
\newcommand{\eeq}{\end{eqnarray}}
\newcommand{\bem}{\begin{multline}}
\newcommand{\eem}{\end{multline}}
\newcommand{\ba}{\begin{align}}
\newcommand{\ea}{\end{align}}

\newcommand{\fp}[2]{\frac{\partial #1}{\partial #2} }

\newcommand{\snorm}[1]{\left|#1\right|}

\renewcommand{\deg}{$^\circ$ }
\newcommand{\pa}{\bfp_\alpha}
\newcommand{\op}{\overline{p}}
\newcommand{\obfp}{\overline{\bfp}}
\newcommand{\obfpa}{\overline{\bfp}_\alpha}
\newcommand{\esa}{\bfeps^s_\alpha}
\newcommand{\opsi}{\overline{\psi}}

\newcommand{\as}{a^{\ast}}
\newcommand{\hpsi}{\hat{\psi}}
\newcommand{\hPsi}{\hat{\Psi}}
\newcommand{\hC}{\hat	{C}}
\newcommand{\hh}{\hat	{h}}
\newcommand{\he}{\hat	{e}}
\newcommand{\hatl}{\hat{l}}
\newcommand{\hG}{\hat{\Gamma}}
\newcommand{\heta}{\hat{\eta}}
\newcommand{\hzeta}{\hat{\zeta}}
\newcommand{\hc}{\hat{c}}
\newcommand{\hchi}{\hat{\chi}}

\newcommand{\cphi}{\check{\varphi}}
\newcommand{\cp}{\check{p}}

\newcommand{\et}{\Tilde{\bfe}}
\newcommand{\tW}{\overset{\sim}{W}}

\DeclareMathOperator{\sgn}{sgn}

\newcommand{\kVcm}{\,\mathrm{kV/cm}}

\begin{document}
\bibliographystyle{elsarticle-harv}

\begin{frontmatter}

\title{A phase-field model for ferroelectrics with general kinetics.\\ Part I: Model formulation}

\author[ep,eth]{Laurent~Guin}
\author[eth]{Dennis~M.~Kochmann\corref{ca}}
\ead{dmk@ethz.ch}
\cortext[ca]{Corresponding author}
\bigskip\bigskip
\address[ep]{LMS, CNRS, \'Ecole polytechnique, Institut Polytechnique de Paris, 91128 Palaiseau, France}
\address[eth]{Mechanics \& Materials Lab, Department of Mechanical and Process Engineering, ETH Zürich, 8092 Zürich, Switzerland}

\begin{abstract}
When subjected to electro-mechanical loading, ferroelectrics see their polarization evolve through the nucleation and evolution of domains.
Existing mesoscale phase-field models for ferroelectrics are typically based on a gradient-descent law for the evolution of the order parameter. While this implicitly assumes that domain walls evolve with linear kinetics, experiments instead indicate that domain wall kinetics is nonlinear. This, in turn, is an important feature for the modeling of rate-dependent effects in polarization switching.
We propose a new multiple-phase-field model for ferroelectrics, which permits domain wall motion with nonlinear kinetics, with applications in other solid-solid phase transformation problems. By means of analytical traveling wave solutions, we characterize the interfacial properties (energy and width) and the interface kinetics of straight domain walls, as furnished by the general kinetics model, and compare them to those of the classical Allen--Cahn model. We show that the proposed model propagates domain walls with arbitrarily chosen nonlinear kinetic relations, which can be tuned to differ for the different types of domain walls in accordance with experimental evidence. 
\end{abstract}

\begin{keyword}
Ferroelectrics; Phase transformation; Phase-field model; Domain wall; Interface kinetics.
\end{keyword}

\end{frontmatter}
%

\section{Introduction\label{sec:intro}}

Ferroelectrics are a prime example of multifunctional materials with electro-mechanical coupling. They are largely used as sensors and actuators for their piezoelectric properties \citep{Uchino2018} as well as in ferroelectric random-access memory \citep{Ishiwara2004}, which uses the reversibility of spontaneous polarization states.  Below a critical temperature, referred to as the Curie temperature, ferroelectric ceramics possess a polar structure with a \emph{spontaneous electric polarization}, which can be reversed under the application of electro-mechanical loading. Polarization switching is accommodated by transformations between lattice variants of different orientations---e.g., the six orientations of the tetragonal variants for barium titanate (BaTiO$_3$) at room temperature---and between phases of distinct crystal symmetry types---e.g., the  tetragonal and rhombohedral phases in some compositions of lead zirconate titanate (PZT). 

At the mesoscale (tens of nanometers to hundreds of microns), polarization reversal occurs through the emergence and evolution of a \emph{microstructure} formed by the different variants (and phases, where applicable) referred to as ferroelectric \emph{domains} separated by \emph{domain walls}. In defect-free ferroelectric single-crystals, the microstructure that develops during switching vanishes at the end of that process, which leads to a single polarization domain in the sample \citep{Little1955}. By contrast, in polycrystalline ferroelectric ceramics, grains with different orientations exert mechanical constraints, so that microstructural arrangements with multiple ferroelectric domains remain after polarization switching \citep{Arlt1980}. 

At the macroscale, a few representative curves are commonly used to characterize polarization switching. First, \emph{bipolar measurements} are obtained by cycling the applied electric field between $\pm e_{\max}$. These yield the $p$-$e$ hysteresis curve and the strain butterfly curve, giving, respectively, the evolution of the average polarization and longitudinal strain (under stress-free conditions) as functions of the applied electric field. When bipolar measurements are performed at a sufficiently low rate (typically with frequency $\omega \lesssim 10^{-3}\,\mathrm{Hz}$), no rate dependence is observed, suggesting that the kinetics of domain evolution only plays a minor role. By contrast, at higher rates, bipolar curves show a strong rate dependence, a consequence of the finite kinetics of domain evolution (see, e.g., \citealt{Wieder1957,Chen2011} and \citealt{Wieder1957,Schmidt1981,Yin2002,Chen2011,Kannan2022} for rate-dependent hysteresis curves in single and polycrystals, respectively). Another type of experiment, referred to as \emph{pulse experiment}, determines the time evolution of the average polarization and strain in response to an applied pulse of the electric field (see, e.g., \citealt{Merz1954,Merz1956,Hubmann2016} for single crystals and \citealt{Genenko2012,Schultheiss2018,Schultheiss2019} for polycrystals). While no significant polarization reversal occurs during the short rise time of pulse loading ($\sim 0.1\, \mu \mathrm{s}$), polarization switching does occur during the subsequent constant applied electric field. Overall, the profile of the time evolution of polarization and strain is a fingerprint at the macroscale of the sequence of switching mechanisms at the mesoscale, namely, the nucleation of domains and their subsequent evolution with finite kinetics. Hence, a proper modeling of the kinetics of domain evolution is crucial for understanding the macroscopic response of ferroelectrics as given by rate-dependent bipolar measurements or pulse experiments. 

At the mesoscale, continuum models  that resolve the time evolution of the domain microstructure together with the local distribution of electric and mechanical fields are of two sorts: (i) \emph{sharp-interface models}, which describe domain walls as surfaces of discontinuity with zero thickness, and (ii) \emph{diffuse-interface models} (also referred to as \emph{phase-field models}), in which the electro-mechanical fields vary steeply but continuously across domain walls.

In sharp-interface models, the classical balance laws of electro-quasistatics and mechanics provide field equations within ferroelectric domains and jump conditions at the domain walls. These are completed by constitutive relations that account for the elastic, dielectric, and piezoelectric properties of the ferroelectric variants. Domain walls are twin interfaces or phase boundaries, for which it is known from the continuum theory of phase transitions  that their velocity is, in the subsonic regime, not uniquely determined by the above system of equations \citep{Abeyaratne2006}. Hence, the jump conditions are complemented by an additional relation termed the \emph{kinetic relation}\footnote{
In the remainder of this article the term \emph{kinetic relation} refers to the relation, defined for sharp-interface models, between the normal velocity of the interface and the conjugate driving traction---the latter being defined as the quantity whose product with the normal velocity represents the dissipation rate \citep{Abeyaratne2006}.
}
between the normal velocity of the interface and the local \emph{driving traction}. This framework, first developed in the mechanics literature within the thermo-mechanical setting \citep{Truskinovsky1987,Abeyaratne1990}, was later extended to include the interaction with electromagnetic fields \citep{Jiang1994a} and applied to ferroelectrics in \citet{Jiang1994b,Rosakis1995,Loge1996,Kessler2006}. Nevertheless, due to the presence of moving surfaces of discontinuity, the sharp-interface model does not lend itself to numerical simulations, which is why in practice its use is limited to analytical works.

In parallel, \emph{diffuse-interface} approaches based on the \emph{time-dependent Ginzburg-Landau} (TDGL) model were developed to simulate the evolution of the patterns formed by ferroelectric domains. Based on  the early modeling of the structure of domains in ferromagnets by \citet{Landau1935}, Ginzburg-Landau potentials---with multiwelled functions of the polarization vector supplemented with polarization gradient terms---were used to model the dependence on temperature of properties of ferroelectric domain walls, such as their thickness and energy, when close to the paralectric-ferroelectric transition temperature \citep{Zhirnov1959,Bulaevskii1964,Cao1991}. 
Later, and initially in the physics and material science literature, these works were extended by the TDGL formulation with the purpose of investigating the paraelectric--ferroelectric transition with a focus on the velocity of the paraelectric--ferroelectric interface \citep{Gordon1986} and the formation and evolution of patterns formed by the different tetragonal variants. The latter arise in  ferroelectrics quenched below the Curie temperature, as a consequence of the paraelectric--ferroelectric transition \citep{Nambu1994,Ahluwalia2001,Yang1995,Hu1997,Hu1998,Li2001}. 
Afterwards, works in the the mechanics literature dealt with the coupling of the TDGL equations---which were re-derived from variational principles \citep{Zhang2005} or microforce balance \citep{Su2007}---with electrostatic Gauss' law and the mechanical balance equations to simulate the evolution of ferroelectric domain patterns in response to an applied electric field, i.e., \emph{polarization switching} \citep{Zhang2005,Zhang2005a,Su2007,Vidyasagar2017,Indergand2020}.

While these later works successfully accounted for features of the quasistatic $p$-$e$-hysteresis and strain butterfly curves, the TDGL model reaches its limitations when it comes to modeling rate effects as well as the material response to pulse switching experiments. For instance, TDGL simulations of polarization switching in response to applied electric field pulses
of magnitude $e_\text{sw}$ yield a switching time $\tau_s$ that scales as $\tau_s \propto 1/e_\text{sw}$ \citep{Indergand2019} in contradiction to the experimental scaling $\tau_s \propto \mathrm{e}^{1/e_\text{sw}}$ \citep{Merz1956,Schultheiss2019}, known as \textit{Merz law}.

The above scaling that results from the TDGL model is expected, since the TDGL formulation postulates a linear relation between the time evolution of the order parameter (polarization $\bfp$) and the (negative of the) variational derivative of the free-energy density $\Psi$, i.e.,
\begin{equation} \label{eq:acIntro}
\mu \dot \bfp =  -\frac{\delta \Psi}{\delta \bfp}
\end{equation}
with the inverse mobility coefficient $\mu>0$. As will be discussed in Section~\ref{sec:ac}, when the TDGL model is viewed as the regularization of a sharp-interface model, the evolution law \eqref{eq:acIntro}, also known as the \textit{Allen--Cahn equation} \citep{Allen1979}, amounts to assuming a linear kinetic relation. As reviewed in~\ref{app:DWdynamics}, this point is precisely in contrast with experimental evidence, which shows that domain wall motion is governed by nonlinear kinetic relations. Hence, modeling time-dependent switching with a diffuse-interface approach requires a new phase-field formulation, one that permits nonlinear domain wall kinetics. This is the objective of the present work, which focuses on the theoretical foundation. Part~II of this study will present numerical applications.

Domain evolution in ferroelectrics is a specific case of the general class of problems of structural phase transformations. These include, e.g., stress-induced phase transformations (such as the martensite--austenite transformation) and deformation twinning. In this more general setting, the question of how to formulate a regularized model that accounts for the nonlinear kinetics of interfaces in solids was considered before. \citet{Hou1999} used a level-set method based on the Hamilton-Jacobi equation \citep{Osher1988} to model interface propagation with nonlinear and anisotropic kinetics. \citet{Alber2013} proposed a phase-field model for two-phase solid-solid transformations---termed the \emph{hybrid model} because it shares properties of a Hamilton-Jacobi and a parabolic equation---which permits nonlinear kinetics \citep[see also][]{Alber2005,Alber2007}. The model proposed in Section~\ref{sec:gkm} of this article may be interpreted as a reformulation of that of \cite{Alber2013}, which is specialized for ferroelectrics and extended to transformations between multiple phases (as opposed to only two in the works of \citet{Alber2005,Alber2007,Alber2013}). On a different front, \citet{Agrawal2015,Agrawal2015a} recently proposed a phase-field model that admits nonlinear kinetics. Their evolution equation closely resembles the one of \citet{Hou1999}, reformulated as a phase-field model instead of a level-set method. However, the model of \citet{Agrawal2015} does not lend itself easily to extensions to multiple phases (see  \cite{Agrawal2016}). Finally, \citet{Tuma2018} included a mixed-type dissipation potential---combining viscous and rate-independent contributions---in the variational formulation of a phase-field model for displacive phase transformations such as martensitic transformations. This furnishes a nonlinear kinetic law in the corresponding sharp-interface model, one that prescribes a threshold on the driving force for interface motion. This model was later extended by including this feature in a multi-phase-field model for solid-solid transformations, which rests upon a micromorphic formulation \citep{Rezaee2021}. While these two formulations introduce in a neat fashion a specific nonlinearity in the kinetic relation (specifically, a threshold on the driving force), our goal here is to include a kinetic relation with general nonlinearity so as to accurately account for experimentally measured kinetics. Within this setting of diffuse-interface models with nonlinear kinetics for solid-solid transformations, the one we introduce in Section~\ref{sec:gkm} is the first to encompass multiple-phase transformations and embed arbitrary nonlinearity. It finds a particularly relevant application in ferroelectric domain evolution, since experimental evidence confirms that the kinetic relation for domain walls is nonlinear. Further, the proposed model allows to select different kinetics for the different types of domain walls (e.g., 90\deg  and 180\deg domain walls in tetragonal ferroelectrics), as it is indicated by experimental results.

The remainder of this article is organized as follows. In Section~\ref{sec:sharp} we introduce a sharp-interface model for ferroelectrics, which constitutes the basis for discussing the characteristics of the diffuse-interface models of subsequent sections. Section~\ref{sec:ac} reviews the characteristics of classical models based on the Allen--Cahn equation with regards to the properties (interfacial energy and thickness) and kinetics of 90\deg  and 180\deg domain walls. In Section~\ref{sec:gkm}, we introduce a new phase-field model for ferroelectrics with general kinetics, and we discuss its properties with focus on the kinetics of the two types of domain walls. The main features of the new phase-field model are discussed in the conclusion in Section~\ref{sec:conc}. Empirical data on the nonlinear kinetics of domain walls in barium titanate coming from the physics literature are briefly reviewed in \ref{app:DWdynamics}. Numerical applications will be reported in Part~II of this study.

\section{The sharp-interface model for ferroelectrics\label{sec:sharp}}

We begin by formulating a mesoscale continuum model for ferroelectrics. Domain walls are represented by sharp interfaces, across which some or all of the electro-mechanical fields exhibit discontinuities. The motion of a domain wall is governed by a function, termed the \emph{kinetic relation}, which relates its normal velocity to the thermodynamic driving traction exerted on the interface. The sharp-interface model introduced in this section serves as a reference  to discuss the behavior of the diffuse-interface models of Sections~\ref{sec:ac} and~\ref{sec:gkm}.

In Section~\ref{sec:gp}, we present the mechanical balance laws, Maxwell's equations and evolution equation for a singular surface, formulated for a general continuum subject to electro-mechanical coupling. In Section~\ref{sec:cr}, we introduce an expression for the electric enthalpy of ferroelectrics, from which  the constitutive relations derive, and we discuss the functional form of the kinetic relation appropriate for domain wall motion in ferroelectrics. The resulting model is specialized to rigid ferroelectrics in Section~\ref{sec:rigidferro}.

Note that we use the following notation of tensor calculus. For any vector fields $\bfv(\bfx,t)$ and $\bfw(\bfx,t)$, second-order tensor fields $\bfR(\bfx,t)$ and $\bfS(\bfx,t)$, and third- and fourth-order tensors $\dsT$ and $\dsU$, respectively, we have in a Cartesian basis
\begin{empheq}[left=\empheqlbrace]{align} 
\begin{alignedat}{4}
&(\nabla \bfv)_{ij}=v_{i,j}, \quad & (\divv \bfR)_i=R_{ij,j}, \quad & (\bfR \bfv)_i= R_{ij}v_j, \quad &(\bfR \cdot \bfS)_{ij}=R_{ik} S_{kj}, \\
 &\bfR : \bfS=R_{ij} S_{ij}, \quad &(\dsT^\mathrm{T} \cdot \bfv)_{ij}= T_{kij} v_k, \quad  &(\dsT: \bfR)_{k}= T_{ijk} R_{jk}, \quad & (\dsU : \bfR)_{ij}= U_{ijkl} R_{kl},
\end{alignedat}
\end{empheq}
where indices following a comma indicate spatial derivatives, and we use Einstein's summation convention.

\subsection{General principles \label{sec:gp}}

We consider a deformable body $\calB$ occupying the region $\Omega \subset \mathbb{R}^3$ in the presence of electrostatic fields filling $\mathbb{R}^3$. The deformation of $\calB$ is described by the displacement field $\bfu(x,t) : \Omega \times \dsR \rightarrow \dsR^3$. Given the small strains encountered in ferroelectric ceramics, we assume linearized kinematics and introduce the infinitesimal strain tensor $\bfeps(\bfx,t)=(\nabla \bfu)_{\mathrm{sym}}=\frac{1}{2}\left[\nabla\bfu+(\nabla\bfu)\T\right]$. In addition, we restrict ourselves to quasistatics both in the mechanical sense (i.e., inertia is neglected) and in the sense of electro-quasistatics (i.e., magnetic induction is neglected) and assume that no free charges exist within $\Omega$. The former is justified by the slow rates of interest, while the latter is a common assumption for undoped ferroelectrics. Let $\calS(t)$ be a regular surface of discontinuity for some or all of the electro-mechanical fields, which separates $\Omega$ into $\Omega_-(t)$ and $\Omega_+(t)$. For  $\bfx \in \calS(t)$ we denote by $\bfn(\bfx,t)$  the unit normal to $\calS(t)$ pointing into $\Omega_+$.

Classically, one derives the field equations in $\Omega \setminus \calS$  and jump conditions on $\calS$  by localizing the integral form of mechanical balances and electrostatic equations on these two sets \citep{Abeyaratne1990,Jiang1994a,Gurtin2009,Kessler2006,Mueller2006}. In the following, we directly use the local forms. Mechanical equilibrium reads
\begin{empheq}[left=\empheqlbrace]{align} \label{eq:mech} 
\begin{aligned}
\divv \bfsigma &= 0 \quad \mbox{in} \quad \Omega \setminus \calS(t), \\
\llbracket \bfsigma \rrbracket \bfn &= 0 \quad \mbox{on} \quad \calS(t),
\end{aligned}
\end{empheq}
where $\bfsigma(\bfx,t)$ denotes the Cauchy stress tensor. For a generic field $w(\bfx,t)$ defined in $\Omega \setminus \calS$ we define the jump  $\llbracket w  (\bfx,t) \rrbracket=w^+(\bfx,t)-w^-(\bfx,t)$, where $w^+(\bfx,t)$ and $w^-(\bfx,t)$ are the limiting values of $w$ at  point $\bfx$ on $\calS(t)$, given by
\begin{equation}
w^+(\bfx,t)=\lim_{\substack{\xi \rightarrow 0 \\ \xi >0}} w(\bfx + \xi \bfn(\bfx,t)) \quad \mbox{and} \quad
w^-(\bfx,t)=\lim_{\substack{\xi \rightarrow 0 \\ \xi <0}} w(\bfx + \xi \bfn(\bfx,t)).
\end{equation}

When it comes to the electrostatic fields, we denote by $\bfe(\bfx,t)$ the electric field, by $\bfd(\bfx,t)$ the electric displacement, and by $\bfp_{mat}(\bfx,t)$ the polarization\footnote{%
Note that we here use the notation $\bfp_\text{mat}$ for the polarization (sometimes referred to as \emph{material polarization}, see e.g., \citet{Schrade2013}). Indeed, we reserve the notation $\bfp$ for what we call the \emph{spontaneous polarization},  i.e., the polarization in a ferroelectric variant at zero stress and electric field (see Section~\ref{sec:cr}).} %
field. All three are defined over $\dsR^3 \times \dsR$ and related by definition through
\begin{equation}
\bfd = \epsilon_0 \bfe + \bfp_\text{mat},
\end{equation}
where $\epsilon_0$ is the permittivity of vacuum. In the absence of free charges, Gauss' law becomes
\begin{empheq}[left=\empheqlbrace]{align} \label{eq:gauss}
\begin{aligned}
\divv \bfd &= 0 \quad \mbox{in} \quad \Omega \setminus \calS(t), \\
\llbracket \bfd \rrbracket \cdot \bfn &= 0 \quad \mbox{on} \quad \calS(t),
\end{aligned}
\end{empheq}
while the Maxwell--Faraday equation in the quasistatic regime yields
\begin{empheq}[left=\empheqlbrace]{align} \label{eq:faraday}
\begin{aligned}
\curl \bfe &= 0 \quad \mbox{in} \quad \Omega \setminus \calS(t), \\
\llbracket \bfe \rrbracket \times \bfn &= 0 \quad \mbox{on} \quad \calS(t).
\end{aligned}
\end{empheq}

In the context of ferroelectrics, surface $\calS$ typically represents domain walls (separating domains with different polarizations) or interphases  (separating different phases, e.g., tetragonal and rhombohedral phases). We use the index $\alpha \in \calD = \{1,\ldots, N\}$ to identify the $N$ different domains or phases. By taking $\bfe$ and $\bfeps$ as the independent state variables, the constitutive relations for $\bfsigma$ and $\bfd$  derive from the electric enthalpy density $W_{\alpha}(\bfe,\bfeps)$ associated with domain $\alpha$ through, respectively,
\begin{equation} \label{eq:cr}
\bfsigma=\frac{\partial W_\alpha}{\partial \bfeps} \quad \mbox{and} \quad \bfd=-\frac{\partial W_\alpha}{\partial e}.
\end{equation}
We associate with the interface $\calS$ 
a constant excess electric enthalpy density $\gamma$, more simply referred to as \emph{interfacial energy}. 
The combination of energy balance and entropy imbalance permits deriving the rate of entropy production $\delta(\bfx,t)$ per unit area of $\calS$ at $\bfx\in \calS(t)$ as 
\begin{equation}
\delta=f V_n,
\end{equation}
where $V_n(\bfx,t)$ is the normal velocity of $\calS$ (taken positive in the direction of $\bfn$), and $f(\bfx,t)$ is the thermodynamic driving traction %
 \citep{Jiang1994a,Mueller2006}\footnote{Note that in \citet{Jiang1994b} the driving traction at the interface is expressed with a different free-energy density, which reads, using our notations, $\tW_\alpha(\bfp_\text{mat},\bfeps)$ with associated constitutive relations $\bfsigma=\partial \tW_\alpha/\partial \bfeps$ and $\bfe=\partial \tW_\alpha/\partial \bfp_\text{mat}$. It is related to the electric enthalpy $W_\alpha(\bfe,\bfeps)$ through $W_\alpha(\bfe,\bfeps)=\tW_\alpha(\bfp_\text{mat},\bfeps)+\frac{\epsilon_0}{2}\bfe \cdot \bfe -\bfe \cdot \bfd$. Rewriting the driving force (4.8) in \citet{Jiang1994b} in terms of $W_\alpha$, one obtains with the help of \eqref{eq:mech}-\eqref{eq:faraday} the first term of \eqref{eq:df1}. Alternatively, this expression has been derived variationally by \citet{Mueller2006}. The term $\gamma \kappa$ in \eqref{eq:df1}, is associated with curvature-driven motion and originates from the fact that $\calS$ is endowed with excess electric enthalpy. Derivations of this classical term can be found, e.g., in  \citet{Gurtin2000} for surfaces in three dimensions or in \citet{Gurtin1993} for curves in two dimensions.}
given by
\begin{equation} \label{eq:df1}
f=\bfn \cdot \llbracket \bfC \rrbracket \bfn + \gamma \kappa.
\end{equation}
In \eqref{eq:df1}, $\kappa(\bfx,t)$ denotes twice the mean curvature of $\calS$ and $\bfC(\bfx,t)$ is the Eshelby tensor provided by 
\begin{equation}
\bfC=W_\alpha \bfI -  \nabla \bfu ^T \cdot \bfsigma + \bfe \otimes \bfd
\end{equation}
in the domain $\alpha$, with $\bfI$ the identity tensor.

Notice from \eqref{eq:df1} that $f$ and $V_n$ are work-conjugate. The motion of $\calS(t)$ is hence specified by a relation between $V_n$ and $f$, called the \emph{kinetic relation}. The latter may be interpreted as an additional constitutive relation, which embeds information about the microscopic processes underlying the motion of $\calS$ and takes the form
\begin{equation} \label{eq:kr1}
V_n=\hat{V}_n (f),
\end{equation}
where $\hat{V}_n\, : \dsR \rightarrow \dsR$ is a function subject to the restriction  
\begin{equation} \label{eq:frestrict}
\hat{V}_n(f)f \geq 0 \quad \mbox{for all} \; f \in \dsR,
\end{equation}
which ensures the positivity of the entropy production rate $\delta$ (required by the second law of thermodynamics). In general, several surfaces of discontinuity between the different domains or phases evolve simultaneously. Interface energy $\gamma$ and kinetic relation $\hat{V}_n(f)$ are characteristic of each interface $\calS$, i.e., they generally depend on the two phases separated by $\calS$.

\subsection{Constitutive laws \label{sec:cr}}

We apply the model presented in Section~\ref{sec:gp} to the case of a ferroelectric ceramic, which exhibits, below the Curie temperature, a tetragonal crystalline structure with $N=6$ variants in three dimensions (3D), as is the case, e.g., for barium titanate (BaTiO$_3$) and some compositions of lead zirconate titanate (PZT). 

\paragraph{Bulk constitutive behavior}

Each variant possesses a spontaneous polarization $\pa$ which, in the orthonormal system $(\et_1,\et_2,\et_3)$ aligned with the principal crystallographic directions, is defined by
\begin{equation} \label{eq:pbasis}
 \bfp_{1}=-\bfp_2= p_0 \et_1, \quad  \bfp_{3}=-\bfp_4= p_0 \et_2 \quad \mbox{and} \quad \bfp_{5}=-\bfp_6=p_0 \et_3,
\end{equation}
where $p_0$ is the magnitude of the spontaneous polarization of the material. Due to the distortion between the cubic and tetragonal phases, each variant exhibits a spontaneous strain $\esa$, which we write as \citep{Kamlah2001}
\begin{equation}
\esa=(\varepsilon_c-\varepsilon_a) \frac{\pa \otimes \pa}{{p_0}^2}+\varepsilon_a \bfI,
\end{equation}
where $\varepsilon_a<0$ and $\varepsilon_c>0$ are, respectively, the spontaneous longitudinal strain along the $a$-axes (orthogonal to the direction of spontaneous polarization) and along the $c$-axis (along the direction of spontaneous polarization)\footnote{
$\varepsilon_a$ and $\varepsilon_c$ can be uniquely defined from the lattice parameters of the tetragonal phase by using, e.g.,  Aizu's definition of spontaneous strain \citep[see][Section~2.1.3]{Tagantsev2010}, which in particular requires volume-preserving spontaneous strain, i.e., $2 \varepsilon_a+\varepsilon_c=0$.}.
With these features, each ferroelectric variant is modeled as a linear elastic, dielectric, and piezoelectric material with the following electric enthalpy:
\begin{equation} \label{eq:eed}
W_\alpha(\bfe,\bfeps)=\underbrace{ - \frac{1}{2} \bfe \cdot \bfepsvar_\alpha \bfe  - \pa \cdot \bfe}_{W_\alpha^{\mathrm{dielec}}(\bfe)} \; + \;\underbrace{\frac{1}{2}(\bfeps-\esa) :\dsC_\alpha : (\bfeps-\esa)}_{W_\alpha^{\mathrm{mech}}(\bfeps)} \; \; \underbrace{- (\bfeps-\esa) : \dsE_\alpha^T \cdot \bfe}_{W_\alpha^{\mathrm{piezo}}(\bfe,\bfeps)} ,
\end{equation}
where $\bfepsvar_\alpha$, $\dsC_\alpha$, and $\dsE_\alpha$  are the second-order dielectric tensor, fourth-order elasticity tensor, and third-order piezoelectric tensor \citep{Standard1949}, respectively, which satisfy the material symmetries of the tetragonal phase. The constitutive relations resulting from \eqref{eq:cr} are
\begin{empheq}[left=\empheqlbrace]{align} \label{eq:cr2}
\begin{aligned}
\bfsigma&=\dsC_\alpha : (\bfeps-\esa)-\dsE_\alpha^T \cdot \bfe,\\
\bfd&=\bfepsvar_\alpha \bfe +\pa  + \dsE_\alpha : (\bfeps-\esa).
\end{aligned}
\end{empheq}

\paragraph{Kinetic relation}

For tetragonal ferroelectrics, we distinguish between two types of domain walls: 180\deg domain walls separating domains with antiparallel polarizations (i.e., interfaces between $\{\alpha,\beta\} \in  \calI_{180}=\{ \{1,2\}, \{3,4\}, \{5,6\} \}$) and 90\deg domains walls, across which polarization fields are orthogonal (these correspond to interfaces between $\{\alpha,\beta\} \in  \calI_{90}= \{ \{ \alpha, \beta \}: \alpha, \beta \in \calD, \alpha \neq \beta \}  \setminus \calI_{180}$). \\

The dynamics of domain walls has been investigated both experimentally and theoretically in bulk single-crystals since the 50's and in epitaxial ferroelectric thin films since 2000. These works, reviewed in \ref{app:DWdynamics}, show that ferroelectrics exhibit two regimes of domain wall dynamics. At low electric fields, the domain wall velocity is described by an inverse exponential law,
\begin{equation} \label{eq:vel1a}
V_n=V_\infty \exp(-e_a/e),
\end{equation}
where $V_\infty$ is a characteristic velocity and $e_a$ an \emph{activation field} which varies as the inverse of temperature. This kinetics has been interpreted as resulting from thermally-activated nucleation and growth of new domains along the domain wall \citep{Miller1960a,Shin2007,Liu2016} or as a creep process of an interface moving in a pinning potential \citep{Tybell2002,Jo2009}. As the electric field increases above a threshold field $e_t$, the wall kinetics experiences a transition to a different regime described by a power law,
\begin{equation}  \label{eq:vel2a}
V_n \propto e^\theta,
\end{equation}
where $\theta$ is an exponent that depends on the dimensionality of the system: $\theta=1.4$ has been reported for bulk BaTiO$_3$ \citep{Stadler1963}, while $\theta=0.7$ was found for PZT epitaxial thin films \citep{Jo2009}. Different theoretical explanations for that second regime have been proposed. On the one hand, in line with the work of \citet{Miller1960a}, this second regime is interpreted in terms of the nucleation of domains at the domain boundary with thicknesses of multiple crystalline unit cells \citep{Stadler1963}. On the other hand, in the picture of an interface moving in a pinning potential, the change in kinetics is seen as a pinning--depinning transition to a non-activated regime, i.e., the wall moves without assistance of thermal fluctuation \citep{Jo2009}. 

\begin{figure}
\centering
\includegraphics[width=0.55\textwidth]{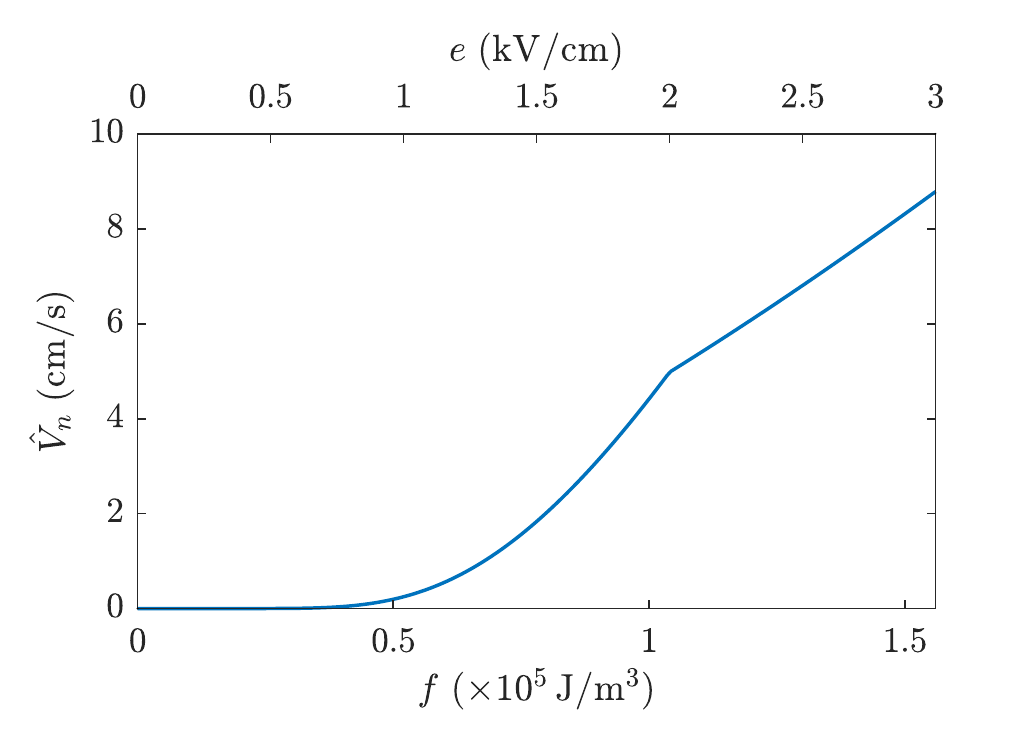}
\caption{Kinetic relation \eqref{eq:kr} with parameters corresponding to 180\deg domain walls in BaTiO$_3$ (see \ref{app:DWdynamics}): $V_l=100\,\mathrm{cm/s}$, $f_a=3 \times 10^5\,\mathrm{J/m^3}$, $f_t=1 \times 10^5\,\mathrm{J/m^3}$, $\theta=1.4$. }
\label{fig:kinrel}
\end{figure}

Relations \eqref{eq:vel1a} and \eqref{eq:vel2a} were obtained on plate-like samples with out-of-plane polarization and electric field. In this configuration, the driving traction reduces to $f= 2 p_0 e$, i.e., it is proportional to the applied electric field~$e$. Therefore, the experimental evidence discussed above indicates that the motion of domain walls is governed by the following kinetic relation:
\begin{equation}\label{eq:kr}
\hat{V}_n(f) = \begin{cases} \displaystyle
 V_l \sgn(f) \exp(-f_a/|f|) \quad &\mbox{for} \quad |f| \leq f_t, \\
 V_h \sgn(f) \left( |f|/f_t \right)^\theta \quad &\mbox{for} \quad |f| > f_t,
\end{cases}
\end{equation}
where $V_l$, $V_h$, $f_a$, $f_t$ and $\theta$ are referred to as the domain wall kinetics parameters and satisfy $V_l \exp(-f_a/f_t) = V_h$ to ensure the continuity of $\hat{V}_n$ at $f_t$. Such a kinetic relation is plotted in Figure~\ref{fig:kinrel} with the parameters corresponding to 180\deg walls in BaTiO$_3$. 
As briefly reviewed in \ref{app:DWdynamics}, experiments indicate that the kinetic relation for 90\deg domain walls differs from that of 180\deg walls either in the coefficients of \eqref{eq:kr} or in the functional form of $\hat V_n (f)$, depending on the ferroelectric material.

\subsection{Specialization to rigid ferroelectrics \label{sec:rigidferro}}

Our goal is a phase-field model that displays the nonlinear kinetics presented in Section~\ref{sec:cr}, in contrast to the linear kinetics that the classical Allen--Cahn equation furnishes. To this end, we make two simplifying assumptions: first, the ferroelectric ceramic is assumed mechanically rigid, so that mechanical couplings are neglected, and, second, the permittivity is assumed isotropic. These assumptions allow us to work in a clean setting, where the new features of the general kinetics model are clearly revealed. As they are not necessarily valid in practice, they will be lifted in future work, when the objective is to provide quantitative predictions.
In this simplified setting, the electric enthalpy density \eqref{eq:eed} of each phase~$\alpha$ reduces to that of a dielectric,
\begin{equation}  \label{eq:eed2}
W_\alpha(\bfe)=- \frac{\epsilon}{2} \bfe \cdot  \bfe  - \pa \cdot \bfe,
\end{equation}
where the scalar permittivity $\epsilon$ is independent of the phase. Consequently, the constitutive relation \eqref{eq:cr2}$_2$ reads
\begin{equation}
\bfd=\epsilon \bfe + \pa.
\end{equation}
In view of \eqref{eq:gauss}$_2$, \eqref{eq:faraday}$_2$ and \eqref{eq:eed}, the driving traction $\eqref{eq:df1}$ is rewritten in the simplified form
\begin{equation} \label{eq:df2}
f=-\langle \bfe \rangle \cdot \llbracket \pa \rrbracket+ \gamma \kappa,
\end{equation}
where $\langle \bfe \rangle=(\bfe_++\bfe_-)/2$.

\section{A phase-field model based on the Allen--Cahn equation\label{sec:ac}}

Given the numerical difficulties related to the implementation of sharp-interface models, the phase-field method has been a method of choice for modeling the evolution of domains in ferroelectric ceramics. Classically, the order parameter is chosen as the material polarization $\bfp_\text{mat}(\bfx,t)$ \citep[see e.g.,][]{Zhang2005,Su2007,Vidyasagar2017} or the spontaneous polarization $\bfp(\bfx,t)$ \citep{Schrade2013,Schrade2014}, and evolves according to the Allen--Cahn equation. 

Anticipating that the phase-field model with general kinetics---introduced in Section~\ref{sec:gkm}---uses $\bfp$ as the order parameter, we introduce a phase-field model based on the Allen--Cahn equation written in terms of $\bfp$, with features of the sharp-interface model of Section~\ref{sec:rigidferro}. It serves as a regularized benchmark model to later clarify the new features of the general kinetics model. In particular, we show that the Allen--Cahn equation yields a linear kinetic relation in the limit of low applied electric fields and that the regularization length has an upper bound for properly approximating the sharp-interface model. 

For notational clarity, we use a hat ($\,\hat{}\,$) to highlight all quantities (regularized energy, domain wall properties, and kinetics) associated with the \emph{Allen--Cahn-based phase-field model} as opposed to their counterparts for the general kinetic model later discussed in Section~\ref{sec:gkm}.

We formulate the model in Section \ref{sec:acform} and characterize the properties (interface width, energy, and kinetics) of 180\deg  and 90\deg domain walls in Section~\ref{sec:1dsol} before a brief summary in Section~\ref{sec:summary}.

\subsection{Model formulation\label{sec:acform}}

\paragraph{A regularized electric enthalpy}

The phase-field model is based on a regularized version $\hPsi(\bfe,\bfp,\nabla \bfp)$ of the electric enthalpy \eqref{eq:eed2}, which induces continuous variations of $\bfp$ across domain walls. We here adopt the electric enthalpy density of \citet{Schrade2013} specialized to rigid ferroelectrics, which is given by
\begin{equation} \label{eq:pfe}
\hPsi(\bfe,\bfp,\nabla \bfp)=\underbrace{-\frac{\epsilon}{2} \bfe \cdot \bfe - \bfp \cdot \bfe}_{W(\bfe,\bfp)} \, + \, \hC_{s} \hpsi_{s}(\bfp) + \frac{\hC_g}{2} |\nabla \bfp |^2,
\end{equation}
where $\hC_{s}$ and $\hC_g$ are constants, and  $\hpsi_{s}(\bfp)$ is a multi-well \emph{separation potential} with minima at the six spontaneous polarization states $\pa$. Letting $\obfp=\bfp/p_0$ be the dimensionless spontaneous polarization, $\hpsi_{s}$ is defined as
\begin{equation} \label{eq:opsis}
\hpsi_{s}(\bfp)=\opsi_s\left(\frac{\bfp}{p_0}\right)
\quad\text{with}\quad
\opsi_s(\obfp)=a_0+a_1\sum_{i=1}^3 \op_i^2+a_2 \sum_{i=1}^3 \op_i^4 + 
a_3 \big(\op_1^2 \op_2^2+ \op_1^2 \op_3^2+\op_2^2 \op_3^2\big) + a_4 \sum_{i=1}^3 \op_i^6
\end{equation}
being a dimensionless sextic polynomial.
As shown in \citet{Schrade2013}, the parameters $\{a_i\}_{i=0\ldots4}$ are uniquely determined by requiring (i)~that the minima of $\opsi_s$ are located at $\{\obfpa\}_{\alpha=1\dots6}$, (ii)~that the 180\deg barrier is normalized, i.e.,
\begin{equation} \label{eq:condmin}
\opsi_s(\et_1)=0, \quad \ \fp{\opsi_s}{\op_1}(\et_1)=0, \quad  \mbox{and} \quad \opsi_s(\bfnull)=1,
\end{equation}
and (iii)~by specifying the relative height $0<\hh_{90}<1$ and location $\hchi$ of the 90\deg barrier through\footnote{%
Note that to ensure $a_1<0$ and $a_4>0$ we have the following constraints: $1-2 \hchi^2 < \hh_{90} < 1-2 \hchi^6$ \citep{Schrade2013}. %
}
\begin{equation} \label{eq:hchi}
\opsi_s\left(\hchi(\et_1+\et_2)\right)=\hh_{90} \quad \mbox{and} \quad \fp{\opsi_s}{\op_1}\left(\hchi(\et_1+\et_2)\right)=0.
\end{equation}
Considering the cubic symmetry of $\opsi_s$, \eqref{eq:condmin} and \eqref{eq:hchi} are sufficient to ensure that conditions (i) and (iii) are satisfied for all six wells and all twelve 90\deg barriers. 
The choice of $\hh_{90}$ and $\hchi$ in relation to the properties of 90\deg domain walls will be discussed in Section~\ref{sec:1dsol}.

\paragraph{Governing equations}

In the regularized phase-field model all fields are continuous over $\calB$, so the differential forms of Gauss' law \eqref{eq:gauss}$_1$ and Faraday's equation \eqref{eq:faraday}$_1$ apply over all $\Omega$. Akin to \eqref{eq:cr}, the electric displacement derives from $\hPsi$ as
\begin{equation} \label{eq:disp}
\bfd=-\fp{\hPsi}{\bfe}=-\fp{W}{\bfe}= \epsilon \bfe + \bfp.
\end{equation}
These equations are supplemented by an evolution law for $\bfp$, which is usually taken as the Allen--Cahn gradient-descent equation:
\begin{equation} \label{eq:ac}
\mu \dot{\bfp}=-\frac{\delta \hPsi}{\delta \bfp}=- \fp{\hPsi}{\bfp}+\divv \left(\fp{\hPsi}{\nabla \bfp} \right).
\end{equation}
In view of \eqref{eq:pfe}, \eqref{eq:ac} becomes
\begin{equation} \label{eq:ac2}
\mu \dot{\bfp}= \bfe - \hC_s \fp{\hpsi_s}{\bfp} +\hC_g \nabla^2 \bfp.
\end{equation}

\paragraph{A numerical characteristic electric field} \label{sec:e0}

In the sharp-interface model, the spontaneous polarization takes the discrete values $\bfp_\alpha$ corresponding to the different domains or phases. In the regularized model, based on the electric enthalpy \eqref{eq:pfe}, the spontaneous polarization $\bfp$ is a continuously varying variable that plays the role of a vector phase field. Away from the interface, the Laplacian term in \eqref{eq:ac2} approximately vanishes and equilibrium values of $\obfp$ are defined by
\begin{equation} \label{eq:equip}
\fp{\opsi_s}{\obfp}(\obfp)=\frac{\bfe}{\hC_s/p_0}.
\end{equation}
Whereas large departures of $\bfp$ from the values $\bfp_\alpha$ are normal in the transition regions of domain walls, the spontaneous polarization $\bfp$ shall remain close to one of the $\bfp_\alpha$ inside the domains. In view of \eqref{eq:equip}, this is the case under the condition that 
\begin{equation} \label{eq:esmall}
| \bfe | \ll \he_c,
\end{equation}
where $\he_c=\hC_s/p_0$ is a numerical characteristic electric field. In practice, \eqref{eq:esmall} is ensured by choosing a sufficiently large $\hC_s$. As we will see in Section~\ref{sec:1dsol}, that is equivalent to selecting a sufficiently small regularization length.

Note that the fulfillment of \eqref{eq:esmall} prevents the nucleation of new domains of polarization from \eqref{eq:ac2}. Therefore, domain nucleation, which is frequently observed in experiments, must be included through an explicit additional condition.

\subsection{Analytical solutions for 180\deg  and 90\deg domain walls \label{sec:1dsol}}

In this section, we compute analytical solutions for straight 180\deg  and 90\deg domain walls, both in equilibrium and in steady-state motion. When adopting the form of traveling waves, these solutions provide relations between the numerical parameters $\hC_s$, $\hC_g$, and $\mu$ of the phase-field model and the properties of domain walls (width, interfacial energy, and kinetic relation).

\subsubsection{Equilibrium profile of a neutral 180\deg domain wall} \label{sec:180dw}

\begin{figure}
\centering
\includegraphics[width=0.87\textwidth]{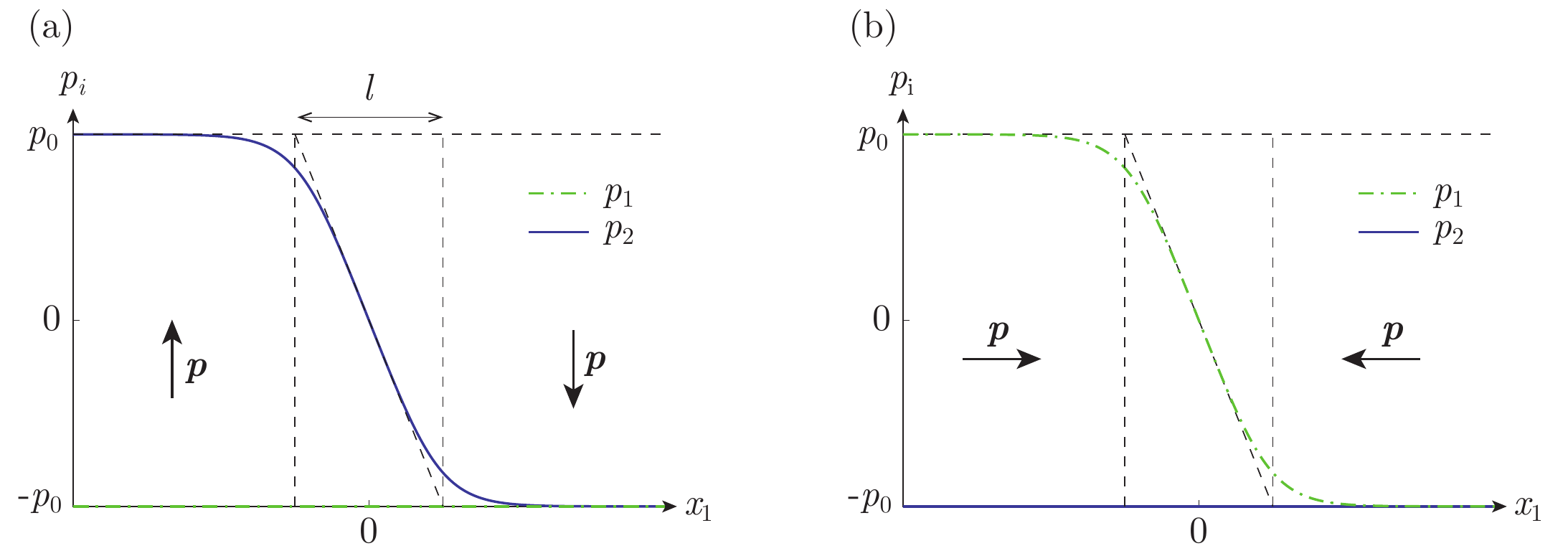}
\caption{ Polarization profiles of two 180\deg domain walls: (a) charge-neutral wall, (b) charged wall.  The profile in (a) corresponds to the solution \eqref{eq:anal180} of \eqref{eq:ode1} for $e=0$, and the profile in (b) is the solution of  \eqref{eq:ode180c} under the equilibrium condition $d=0$.}
\label{fig:profile180}
\end{figure}

We consider the one-dimensional (1D) case of a 180\deg domain wall with spontaneous polarization $\bfp(\bfx,t)=p_2(x_1,t) \et_2$ varying from the equilibrium close to $p_0 \et_2$ at $x_1 \rightarrow - \infty$ to the one near $-p_0 \et_2$ at $x_1 \rightarrow + \infty$, as schematically shown in Figure~\ref{fig:profile180}(a).
This configuration is referred to as \emph{neutral} 180\deg domain wall due to the continuity of the normal component of $\bfp$ across the wall, which implies the absence of bound charges in the wall.  In view of \eqref{eq:disp} and noting that the polarization profile is divergence-free, we can assume a uniform electric field $\bfe=e \et_2$ while satisfying Gauss' law \eqref{eq:gauss}$_1$.

We look for a traveling wave solution of the form
\begin{equation} \label{eq:tr180}
p_1(x_1,t) = 0,
\quad
p_2(x_1,t)=p_0 \cp_2(x_1-v_{180} t),
\end{equation}
where $\xi=x_1-v_{180} t$ is the co-moving coordinate and $v_{180}$ the velocity of the wall. We denote by $\hpsi_{180}(\ol{p}_2)=\opsi_s(\ol{p}_2\et_2)$ the section of the separation potential relevant for  this situation. Inserting \eqref{eq:tr180} into \eqref{eq:ac2} yields an ordinary differential equation for $\cp_2(\xi)$:
\begin{empheq}[left=\empheqlbrace]{align} \label{eq:ode1}
\begin{aligned}
\hC_g p_0 \cp_{2,\xi \xi} + \mu v_{180} p_0 \cp_{2,\xi}-\frac{\hC_s}{p_0} \hpsi'_{180}(\cp_2)&=-e,\\
\hpsi'_{180}(\cp_2(\pm\infty))&=e p_0/\hC_s,
\end{aligned}
\end{empheq}
where \eqref{eq:ode1}$_2$ are the far-field boundary conditions at $\pm \infty$, and the prime denotes the derivative of a single-variable function with respect to its variable.

\paragraph{The equilibrium profile: $e=0$}

In the absence of an electric field, \eqref{eq:ode1} admits a static analytical solution with $v_{180}=0$ \citep{Cao1991}:
\begin{equation} \label{eq:anal180}
\cp_2^\text{eq}(\xi)=\frac{\sinh(\xi/\hat \lambda_{180})}{\sqrt{\hat A+\sinh^2(\xi/\hat \lambda_{180})}}, \quad \mbox{with}
\quad 
\hat A=\frac{3 a_4 + a_2}{2 a_4 +a_2} \quad \mbox{and} \quad \hat \lambda_{180}=p_0 \sqrt{\frac{\hC_g}{\hC_s(6a_4 + 2 a_2)}}.
\end{equation}
Based on this equilibrium profile, we relate $\hC_s$ and $\hC_g$ to the interface width and energy.
We define the width $\hatl$ of the 180\deg domain wall (see Figure~\ref{fig:profile180}) as
\begin{equation} \label{eq:ldef}
\hatl={2}/{ \cp_{2,\xi}^\text{eq}(0)}
\end{equation}
and its interfacial energy as
\begin{equation} \label{eq:gamma}
\hG=\int_{-\infty}^{+\infty} \hPsi(\bfe,\check \bfp^\text{eq},\nabla \check \bfp^\text{eq}) \dd \xi-\int_{-\infty}^{+\infty} W(\bfe,\bfp^\text{sharp}) \dd \xi,
\end{equation}
where
\begin{equation}
\bfp^\text{sharp}(\xi)=\begin{cases}
+ p_0 \et_2 \quad \mbox{for} \quad \xi < 0, \\
- p_0 \et_2 \quad \mbox{for} \quad \xi \geq 0
\end{cases}
\end{equation}
is the sharp-interface solution of the 180\deg domain wall. Noting that $\bfe$ is uniform and equal in both the phase-field and sharp-interface formulations and that the profile $\check \bfp^\text{eq}(\xi)$ is symmetric, \eqref{eq:gamma} reduces to
\begin{equation} \label{eq:gamma2}
\hG=\int_{-\infty}^{+\infty} \left( \hC_s \hpsi_s(\check \bfp^\text{eq}) + \frac{\hC_g}{2} | \nabla \check \bfp^\text{eq} |^2  \right) \dd \xi=\int_{-\infty}^{+\infty} \left(\hC_s \hpsi_{180} \big(\cp_2^\text{eq} (\xi) \big)  + \frac{\hC_g p_0^2 \cp_{2,\xi}^\text{eq}(\xi)^2}{2}\right) \dd \xi. 
\end{equation}
The first integral of \eqref{eq:ode1} in the equilibrium case ($e=0$, $v_{180}=0$), obtained by multiplying $\eqref{eq:ode1}$ by $\cp_{2,\xi}^\text{eq}$ and integrating from $\xi=-\infty$ to some arbitrary $\xi$, yields
\begin{equation} \label{eq:equipart}
2 \hC_s \hpsi_{180} \big(\cp_2^{eq} (\xi) \big) ={\hC_g p_0^2 \big(\cp_{2,\xi}^{eq}(\xi)\big)^2}.
\end{equation}
From \eqref{eq:equipart} we extract $\cp_{2,\xi}^{eq}(\xi)=\sqrt{2 \hC_s \hpsi_{180} \big(\cp_2^{eq} (\xi) \big)/(\hC_g p_0^2)}$, which we exploit for the change of variable $\xi \rightarrow \ol{p}_2$ in \eqref{eq:gamma2}. This yields
\begin{equation} \label{eq:gamma3}
\hG=\int_{-\infty}^{+\infty}2 \hC_s \hpsi_{180} \big(\cp_2^{eq} (\xi) \big)  \dd \xi = \sqrt{2 \hC_s \hC_g} p_0 \heta, 
\end{equation}
where $\heta= \int_{-1}^1 \hpsi_{180}(\op) \dd \op= \int_{-1}^1 \sqrt{a_0+a_1 \op^2 + a_2 \op^4 + a_4 \op^6} \dd \op $ is a numerical coefficient.
Further, combining \eqref{eq:equipart} and \eqref{eq:ldef} lets us express the width $\hatl$ as
\begin{equation} \label{eq:l2}
\hatl=p_0 \sqrt{{2\hC_g}/{\hC_s}}.
\end{equation}
Conversely,  \eqref{eq:gamma3} and \eqref{eq:l2} allows us to determine $\hC_g$ and $\hC_s$ from $\hG$ and $\hatl$ via
\begin{equation} \label{eq:cgcs}
\hC_g=\frac{\hG \hatl}{2\heta p_0^2} \quad \mbox{and} \quad \hC_s=\frac{\hG}{\heta \hatl}.
\end{equation}
We note here that the interfacial energy $\hG$ shall be taken as the \emph{physical interfacial energy} $\gamma$ introduced in the sharp interface model, while $\hatl$ is viewed as a \emph{numerical parameter}. Indeed, we take the perspective that the objective of the phase-field model is to properly account for the evolution of the pattern formed by domains or phases. In particular, the accuracy of the polarization profile within the diffuse interface (as compared to the physical profile) is of minor importance; instead, we aim to accurately capture the time evolution of the trace of domain walls. An accurate value of $\hG$ to represent the interfacial energy is important to the extent that it is expected from the sharp interface model (see \eqref{eq:df1} and \eqref{eq:kr}) to affect the evolution of a curved interface. This latter point can be confirmed numerically. 
By contrast, $\hatl$ does not directly affect the evolution of the domain wall, but its choice results from a compromise between the following points:
\begin{itemize}
\item
Because a diffuse interface needs sufficient numerical resolution (e.g., it should be discretized with typically no less than three points across its width in FFT-based schemes \citep{Vidyasagar2017}), $\hatl$ cannot be taken too small.
\item
On the other hand, $\hatl$ cannot be taken too large either, because the characteristic electric field $\he_c=\hC_s/p_0=\hG/(\heta \hatl  p_0)$ introduced in Section~\ref{sec:e0} must be sufficiently large for \eqref{eq:esmall} to be satisfied with the physical electric field (notably to ensure that $\bfp$ remains close to one of the $\bfp_\alpha$ inside domains).
\end{itemize}

\subsubsection{Equilibrium profile of a charged 180\deg domain wall} \label{sec:180dwCharged}

\begin{figure}[b]
\centering
\includegraphics[width=0.35\textwidth]{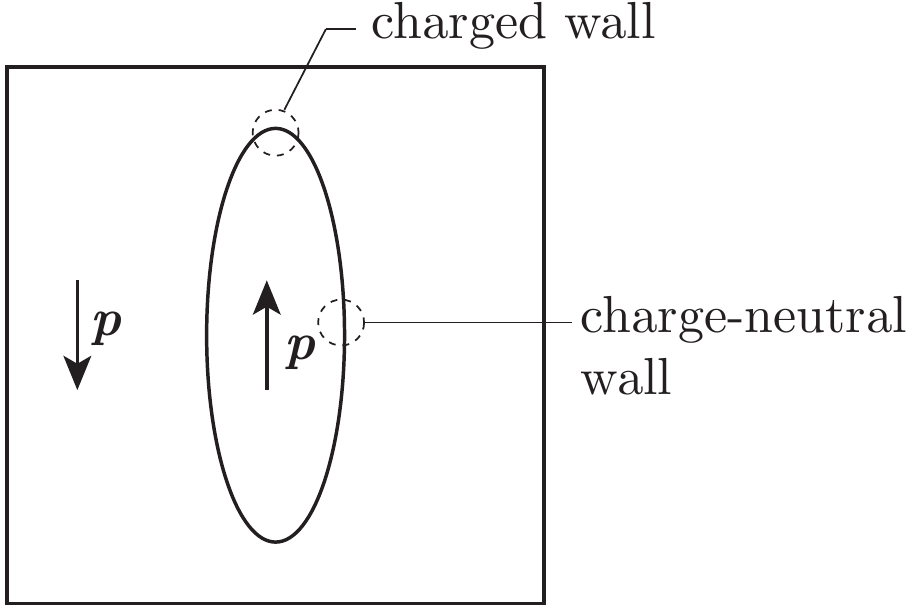}
\caption{Sketch of an elliptical 180\deg inclusion featuring both charge-neutral and charged domain walls.}
\label{fig:elliptic}
\end{figure}

The other characteristic 180\deg domain wall is the so-called \textit{charged} wall, in which the polarization is orthogonal to the domain wall (head-to-head or tail-to-tail). This configuration is usually not considered in the literature \citep[see e.g.][]{Cao1991,Schrade2013,Flaschel2020}, because in the absence of free charges it is associated with high electric fields (of the order of $p_0/\epsilon$). As such, it implies high energies (see \eqref{eq:pfe}) and is hence considered unstable in general. While it is true that ferroelectric domains evolve in such a way as to minimize the existence of charged 180\deg domain walls, these cannot be completely disregarded. Indeed, charged portions of domain walls exist if one considers, e.g., two domains of antiparallel polarization separated by a closed boundary (see Figure~\ref{fig:elliptic}).
This is the natural scenario of a newly nucleated domain in its parent domain.

Let us consider the 1D configuration of a charged 180\deg domain wall with spontaneous polarization $\bfp(\bfx,t)=p_1(x_1,t) \et_1$ varying from the equilibrium close to $p_0 \et_1$ at $x_1 \rightarrow - \infty$ to the one near $-p_0 \et_1$ at $x_1 \rightarrow + \infty$ (see Figure~\ref{fig:profile180}(b)). We assume that the applied electric field is zero in the $\et_2$-direction, which implies, $\bfe(x_1,t)=e_1(x_1,t) \et_1$. As a result, the electric displacement is along the $\et_1$-direction and, because of  \eqref{eq:gauss}$_1$, it is uniform.  Assuming further that it is time-independent, we write $\bfd(x_1,t)=d \et_1$ where $d$ is constant. This allows us to write the non-zero electric field component as
\begin{equation} \label{eq:e180c}
e_1(x_1,t)=\frac{d- p_1(x_1,t)}{\epsilon}.
\end{equation}
We look for a traveling wave solution of the form
\begin{equation} \label{eq:tr180c}
p_1(x_1,t)=p_0 \cp_1(x_1-v_{180}^c t),
\end{equation}
where  $\xi=x_1-v_{180}^c t$ is the co-moving coordinate, $v_{180}^c$ the velocity of the charged wall, and $\cp_1(\xi)$ is a smooth, differentiable function. Inserting  \eqref{eq:tr180c} into \eqref{eq:ac2} and resorting to \eqref{eq:e180c} yields the ordinary differential equation for $\cp_1(\xi)$,
\begin{empheq}[left=\empheqlbrace]{align} \label{eq:ode180c}
\begin{aligned}
\hC_g p_0 \cp_{1,\xi \xi} + \mu v_{180}^c p_0 \cp_{1,\xi}-\frac{\hC_s}{p_0}  {\hpsi^c_{180}}'(\cp_1)&=-\frac{d}{\epsilon},\\
{\hpsi^c_{180}}'\big(\cp_1(\pm\infty)\big)&=\frac{p_0 d}{\epsilon \hC_s},
\end{aligned}
\end{empheq}
where we have introduced the effective double well potential $\hpsi_{180}^c$ for charged walls as
\begin{equation}
\hpsi_{180}^c(\op)=\hpsi_{180}(\op)+\frac{p_0^2}{2 \epsilon \hC_s} \op^2=\hpsi_{180}(\op)+\frac{\heta \hatl p_0^2}{2 \epsilon \hG} \op^2
\end{equation}
with $\hpsi_{180}(\op)$ the cut of $\opsi_s(\obfp)$ defined in Section~\ref{sec:180dw}.

\paragraph{The equilibrium profile: $d=0$}

In the absence of an electric displacement, \eqref{eq:ode180c} admits a static solution with $v_{180}^c=0$. In this case, \eqref{eq:ode180c} differs from \eqref{eq:ode1} only through the effective double-well potential $\hpsi_{180}^c$, which replaces $\hpsi_{180}$. The fact that, physically, the values of $\cp_1$ at infinity given by \eqref{eq:ode180c}$_2$ shall remain close to 1 requires that $\hpsi_{180}^c$ differs little from $\hpsi_{180}$, i.e.,
\begin{equation} \label{eq:fractionSmall}
\hzeta=\frac{\heta \hatl p_0^2}{\epsilon \hG} \ll 1.
\end{equation}
Equation \eqref{eq:fractionSmall} can be seen as an upper bound on $\hatl$, and it is nothing but condition \eqref{eq:esmall} with $\snorm{\bfe} \sim p_0/\epsilon$, which corresponds to the typical electric field that develops in charged domain walls.

One can show that under the assumption \eqref{eq:fractionSmall} the width and interfacial energy of a charged 180\deg domain wall are approximately equal to those of the corresponding neutral walls given by \eqref{eq:gamma3} and \eqref{eq:l2}, respectively. In other words, width and interfacial energy of a 180\deg domain wall are essentially independent of the relative orientation of the wall and the polarization directions.

\subsubsection{Velocity of 180\deg domain walls}

We proceed to consider the steady-state velocity of 180\deg domain walls, beginning with neutral ones and later concluding the solution for charged ones. Taking the first integral of \eqref{eq:ode1}$_1$, we express the velocity $v_{180}$ of the domain wall as
\begin{equation} \label{eq:vel1}
v_{180}=\frac{\hC_s \big[ \hpsi_{180} \big(\cp_2(+\infty)\big) -  \hpsi_{180} \big(\cp_2(- \infty) \big)\big] -e \big[ \cp_2(+\infty)-\cp_2(- \infty) \big]   }{ \mu p_0 \int_{- \infty}^{+ \infty} (\cp_{2,\xi})^2 \dd \xi },
\end{equation}
which involves the unknown function $\cp_{2}(\xi)$. Under the assumption that \eqref{eq:esmall} is satisfied,  the traveling wave problem \eqref{eq:ode1} with $e \neq 0$ is a perturbation of the case  $e=0$. Hence, to first order in $e/\he_c$ the velocity \eqref{eq:vel1} can be approximated using the equilibrium profile $\cp_2^{eq}$, which yields
\begin{equation} \label{eq:vel2}
v_{180} \approx \frac{2 e }{\mu p_0  \int_{- \infty}^{+ \infty} (\cp_{2,\xi}^{eq})^2 \dd \xi  } = \hc_{180} f_{180} ,
\end{equation}
where $\hc_{180}= \hatl/(2 \heta \mu p_0^2)$ is a kinetic coefficient, and $f_{180}=2 e p_0$ denotes the value of the driving traction \eqref{eq:df2} for a straight, neutral 180\deg domain wall.\footnote{For an analogous derivation with mechanical coupling accounted for, see \citet{Indergand2020}.} (In deriving \eqref{eq:vel2}, we used \eqref{eq:equipart} and performed the same change of variable in the integral as in \eqref{eq:gamma3}.) For charged domain walls, the driving traction \eqref{eq:df2} reads $f_{180}^c=2p_0 d/ \epsilon$. One can show that, under the assumption \eqref{eq:fractionSmall}, the linear kinetic relation \eqref{eq:vel2} holds for charged domain walls as well, where it reads $v_{180}^c \approx \hc_{180} f_{180}^c$. We point out that the above analysis, of course, presumes that stable motion of a domain wall at constant speed is feasible. While this is realistic for neutral 180\deg domain walls, the high energy associated with their charged counterparts precludes their stable motion over significant times in general.

\subsubsection{Equilibrium profile of a neutral 90\deg domain wall} \label{sec:90dw}

\begin{figure}
\centering
\includegraphics[width=0.87\textwidth]{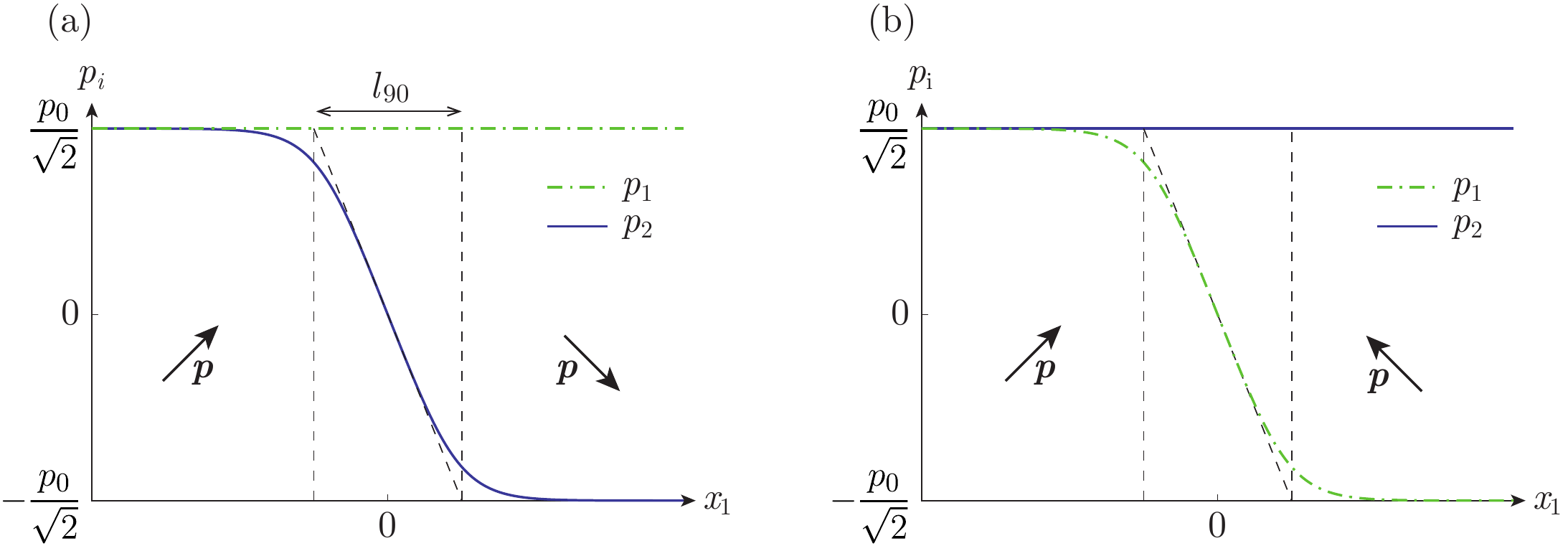}
\caption{Polarization profiles of two 90\deg domain walls: (a) charge-neutral wall, (b) charged wall.  The profile in (a) corresponds to the solution of \eqref{eq:ode2} for $e=0$, while that in (b) is for the equilibrium condition $d=0$.}
\label{fig:profile90}
\end{figure}

For studying 90\deg domain walls, the basis of spontaneous polarizations $\{ \bfp_\alpha \}_{\alpha \in (1,6) }$ defined in \eqref{eq:pbasis} and the multi-well potential \eqref{eq:opsis} are rotated by $+\pi/4$ around the direction $\et_3$.
We consider a 1D head-to-tail 90\deg domain wall with spontaneous polarization $\bfp(\bfx,t)=p_2(x_1,t)\et_2+p_1(x_1,t) \et_1$, varying from the equilibrium close to $p_0 (\et_1+\et_2)/\sqrt{2}$ at $x_1 \rightarrow  - \infty$ to the one near  $p_0 (\et_1-\et_2)/\sqrt{2}$ at $x_1 \rightarrow  + \infty$ (see Figure~\ref{fig:profile90}(a)).
When looking for an equilibrium or a traveling wave solution of \eqref{eq:ac2}, the $p_1$-component appears to vary across the domain wall. As a result, for Gauss' law \eqref{eq:gauss}$_1$ to be satisfied, the electric field is non-uniform and the polarization profile does not admit an analytical solution. 

Therefore, we compute an approximation of the 90\deg domain wall profile, by assuming that $p_1=p_0/\sqrt{2}$ remains constant across the interface. This assumption is all the more valid as the 90\deg barrier in $\opsi_s$ is located along the straight line connecting two orthogonal polarization states, i.e., $\hchi=0.5$ in \eqref{eq:hchi}. As mentioned in Section~\ref{sec:180dw}, in the phase-field formulation we do not aim for an accurate representation of the polarization profile across domain walls, which justifies considering $\hchi$ as a numerical parameter that may be freely chosen. Under this assumption, we suppose that we apply a uniform electric field $\bfe=e \et_2$ while satisfying Gauss' law \eqref{eq:gauss}$_1$. Akin to the 180\deg domain wall (cf.~\eqref{eq:tr180}), we seek a traveling wave solution $p_2(x_1,t)$. Inserting \eqref{eq:tr180}$_2$ in \eqref{eq:ac2}, where $\hpsi_s$ has been properly rotated, yields the following ordinary differential equation for $\cp_2(\xi)$:
\begin{empheq}[left=\empheqlbrace]{align} \label{eq:ode2}
\begin{aligned}
\hC_g p_0 \cp_{2,\xi \xi} + \mu v_{90} p_0 \cp_{2,\xi}-\frac{\hC_s}{p_0} \hpsi'_{90}(\cp_2)&=-e,\\
\hpsi'_{90}\big(\cp_2(\pm\infty)\big)&=e p_0/\hC_s,
\end{aligned}
\end{empheq}
where $v_{90}$ denotes the corresponding velocity. The only difference to \eqref{eq:ode1} lies in the section $\hpsi_{90}(\op_2)$ of $\opsi_s(\obfp)$, given by
\begin{equation}
\hpsi_{90}(\op_2)=\opsi_s\left(\frac{1/\sqrt{2}-\op_2}{\sqrt{2}},\frac{1/\sqrt{2}+\op_2}{\sqrt{2}},0\right)=\as_0+\as_1 \op_2^2+ \as_2 \op_2^4+\as_4 \op_2^6,
\end{equation}
where
\begin{empheq}[left=\empheqlbrace]{align} 
\begin{aligned}
\as_0&=a_0+a_1/2+a_2/8+a_3/16+a_4/32, \qquad \as_1=a_1+3a_2/2-a_3/4+30a_4/32,\\
\as_2&=a_2/2+a_3/4+60a_4/32, \qquad \mbox{and} \quad \as_4=a_4/4.
\end{aligned}
\end{empheq}
Here and in the following, superscript $^\ast$ indicates quantities associated with 90\deg domain walls.

\paragraph{The equilibrium profile: $e=0$}

For $e=0$, \eqref{eq:ode2} admits an analytical solution $\cp_2^{\text{eq}\ast}(\xi)$ similar to \eqref{eq:anal180}. Noting that  $\cp_2^{\text{eq}\ast}(\xi)$ varies from $-1/\sqrt{2}$ at $\xi \rightarrow -\infty$ to $1/\sqrt{2}$ at $\xi \rightarrow \infty$, we define the width of the 90\deg domain wall as
\begin{equation}
\hatl_{90}=\frac{\sqrt{2}}{\cp_2^{\text{eq} \ast}(0)}.
\end{equation}
The interfacial energy $\hG_{90}$ has the same definition as \eqref{eq:gamma}, where now $\check \bfp^\text{eq}$ and $\bfp^\text{sharp}$ are replaced by the phase-field and sharp-interface profiles of the 90\deg domain wall, respectively. The derivation \eqref{eq:gamma2}-\eqref{eq:gamma3} holds with $\hpsi_{90}(\op_2)$ in place of $\hpsi_{180}(\op_2)$, and we obtain
\begin{equation} \label{eq:gamma90}
\hG_{90}=\frac{\heta^*}{\heta} \hG,
\end{equation}
with $\heta^*=\int_{-1/\sqrt{2}}^{1/\sqrt{2}} \hpsi_{90}(\op_2) \dd \op=\int_{-1/\sqrt{2}}^{1/\sqrt{2}} \sqrt{a_0^*+a_1^* \op^2 + a_2^* \op^4 + a_4^* \op^6} \dd \op $. In addition, re-writing \eqref{eq:equipart} for the 90\deg domain wall lets us compute $\cp_{2,\xi}^{\text{eq} \ast}(\xi)$ and derive
\begin{equation} \label{eq:l90}
\hatl_{90}=\frac{\hatl}{\sqrt{2 \hh_{90}}}.
\end{equation}

\subsubsection{Equilibrium profile of a charged 90\deg domain wall} \label{sec:c90dw}

The charged 90\deg domain wall corresponds to the head-to-head (or tail-to-tail) configuration (see Figure~\ref{fig:profile90}(b)). The transition from neutral to charged 90\deg domain walls is in all points equivalent to the transition from neutral to charged 180\deg domain walls. To summarize the main point, for charged 90\deg domain walls an effective potential
\begin{equation}
\hpsi_{90}^c(p_1)=\hpsi_{90}(p_1)+\frac{\heta \hatl p_0^2}{2 \epsilon \hG} p_1^2
\end{equation}
appears in the traveling wave solution (as in \eqref{eq:ode180c}). Likewise, the width and interfacial energy of the charged 90\deg -domain wall are approximately equal to those of the neutral wall under condition \eqref{eq:fractionSmall}. 

\subsubsection{Velocity of 90\deg domain walls}

Following the same procedure as for 180\deg domain walls, the first integral of \eqref{eq:ode2} allows us to write $v_{90}$ as an expression analogous to \eqref{eq:vel1}, where one substitutes $\hpsi_{90}$ for $\hpsi_{180}$. In the regime $e \ll \he_c$, $\cp_2$ can be approximated by the equilibrium profile of \eqref{eq:ode2}, which yields
\begin{equation} \label{eq:kin90}
v_{90} \approx \frac{\sqrt{2}e}{\mu p_0 \int_{-\infty}^{+\infty} (\cp_{2,\xi}^{\text{eq}*})^2 \dd \xi }= \hc_{90} f_{90} ,
\end{equation}
where $\hc_{90}=\hatl/(2\heta^* \mu p_0^2 )$ and $ f_{90}=  \sqrt{2} e p_0$ denote, respectively, the kinetic coefficient and driving traction associated with the 90\deg domain wall. Under the condition \eqref{eq:fractionSmall}, the charged 90\deg domain wall shows the same linear kinetics, with coefficient $\hc_{90}$, as its neutral counterpart.

\subsection{Summary of the properties of 180\deg  and 90\deg domain walls\label{sec:summary}}

\paragraph{Interface energy and width}

We have computed exactly the interfacial energy~$\hG$ and width~$\hatl$ of a neutral 180\deg domain wall and found that these can be arbitrarily prescribed by choosing $\hC_g$ and $\hC_s$ according to \eqref{eq:cgcs}. While we expect to use the physical domain wall energy for $\hG$, $\hatl$ is viewed as a numerical parameter to be chosen in light of considerations discussed above (the more stringent condition on $\hatl$ being \eqref{eq:fractionSmall}). The charged 180\deg domain wall has approximately the same equilibrium properties as the neutral one, as long as \eqref{eq:fractionSmall} is satisfied.

For the classical head-to-tail 90\deg domain wall, we have derived an approximation of its interfacial energy $\hG_{90}$ and width~$\hatl_{90}$, whose accuracy is best for $\hchi=0.5$.  
The ratio $\hG_{90}/\hG$ can be set according to its physical value by independently setting $\hh_{90}$. Indeed, \eqref{eq:gamma90} indicates that this ratio is equal to that of the integrals over the 90\deg  and 180\deg barriers in the multi-well potential $\opsi_s(\obfp)$. Finally, the width~$\hatl_{90}$ is automatically determined by \eqref{eq:l90} and cannot be set independently. Note that, because we view $\hatl_{90}$ as a numerical regularization length, there is no need to prescribe it independently. However, where it is viewed as the physical width of a 90\deg domain wall, the phase-field framework can be modified such that $\hatl_{90}$ is chosen in agreement with its physical value \citep{Flaschel2020}.

\paragraph{Kinetics}

We have found that, in the regime $\snorm{\bfe} \ll \he_c$, the Allen--Cahn evolution equation \eqref{eq:ac} implies approximately a linear kinetic relation between the velocity of domain walls and the driving traction. In addition, the 180\deg  and 90\deg domain walls have different kinetic coefficients related by the ratio $\heta^*/\heta$. Whereas the choice of inverse mobility $\mu$ allows us to independently set one kinetic coefficient, the second one is automatically determined by the ratio of interfacial energies (see \eqref{eq:gamma90} and \eqref{eq:kin90}).

This linear kinetics is not representative of the complex nonlinear kinetics of domain walls, which experiments and atomic-scale modeling discussed in Section~\ref{sec:cr} show. Yet, a proper account of the kinetics of domain walls is important in two respects. First, different kinetics yield different microstructure evolution\footnote{Consider, e.g., how an initially ellipsoidal nucleus evolves, keeping in mind that the driving traction is nonuniform over the surface of the nucleus.}, hence a proper account of the kinetics is needed for obtaining an accurate description of switching at the mesoscale. Second and more importantly, rate-dependent effects in the macroscopic footprints of polarization switching---such as those discussed in Section~\ref{sec:intro}---are a direct consequence of the kinetics of domain walls.
Having a phase-field model that permits domain wall motion with nonlinear kinetics requires us to revise the evolution equation \eqref{eq:ac}. This is the subject of the general kinetics model introduced in Section~\ref{sec:gkm}.

\section{General kinetics model \label{sec:gkm}}

In this section, we introduce a new phase-field model for ferroelectrics, which regularizes the sharp-interface model of Section~\ref{sec:sharp} while conserving nonlinear and independent kinetics for 180\deg  and 90\deg domain walls. We first formulate the model in Section~\ref{sec:gkmform}, before computing analytical traveling wave solutions for 180\deg  and 90\deg domain walls in Section~\ref{sec:1dsolgkm}. The main properties of straight domain walls are summarized and compared to those of the Allen--Cahn model in Section~\ref{sec:gkmsum}. A numerical implementation of the present model will be performed in Part~II, which must also account for the special behavior of curved domain walls and triple and quadruple points at which multiple phases meet.

\subsection{Model formulation\label{sec:gkmform}}

\paragraph{A regularized electric enthalpy}

In the general kinetics model, we use a multi-phase field $\bfvarphi(\bfx,t) \in (0,1)^N$, where $N$ is the number of domains, and for tetragonal ferroelectrics we use $N=6$. Let $\varphi_\alpha$ denote the volume fraction of domain $\alpha$, so that the spontaneous polarization is defined as
\begin{equation} \label{eq:phat}
\bfp(\bfx,t)=\tilde \bfp \big(\bfvarphi(\bfx,t)\big)=\sum_{\alpha=1}^6 \varphi_{\alpha}(\bfx,t) \bfp_{\alpha}.
\end{equation}
We introduce the electric enthalpy ${\Psi}$, which regularizes $W(\bfe,\bfp)$, as
\begin{equation} \label{eq:psihat}
\Psi(\bfe,\bfvarphi,\nabla \bfvarphi)=\underbrace{-\frac{\epsilon}{2} \bfe \cdot \bfe - \Big( \sum_{\alpha=1}^6 \varphi_\alpha \bfp_\alpha \Big) \cdot \bfe}_{W(\bfe, \tilde \bfp(\bfvarphi))} \, + \, C_{s} \psi_{s}(\bfvarphi) + p_0^2 C_g \sum_{\alpha=1}^6 |\nabla \varphi_\alpha |^2,
\end{equation}
where $C_s$ and $C_g$ are two positive constants, and $\psi_s$ is a multi-well potential with minima at (or in the neighborhood of) the six values of $\bfvarphi$ associated with the spontaneous polarization states: $\bfvarphi=(1,0,\ldots,0)$ and its permutations. There exists several possibilities for the choice of $\psi_s$, and the optimal one depends on the values of the physical parameters. Here, we discuss two simple forms for $\psi_s$ that admit analytical insight, even though the framework is sufficiently general to allow for other choices as well.
For analytical derivations, we introduce
\begin{equation} \label{eq:hpsis}
\psi_s^{(n)}(\bfvarphi)=\omega^{(n)}
(\bfvarphi)+\tau(\bfvarphi),
\end{equation}
with
\begin{empheq}[left=\empheqlbrace]{align} \label{eq:omtau}
\begin{aligned}
    \omega^{(n)}
(\bfvarphi)&=4^n \Bigg( \sum_{\{\alpha,\beta \} \in \calI_{180} } \snorm{\varphi_\alpha}^n \snorm{\varphi_\beta}^n + h_{90}  \sum_{\{\alpha,\beta \} \in \calI_{90} } \snorm{\varphi_\alpha}^n \snorm{\varphi_\beta}^n  \Bigg),\\
    \tau(\bfvarphi)&= C_t \sum_{\{\alpha,\beta,\gamma \} \in \calT } \snorm{\varphi_\alpha} \snorm{\varphi_\beta} \snorm{\varphi_\gamma},
\end{aligned}
\end{empheq}
where $h_{90}$ denotes the height of the 90\deg barrier relative to the 180\deg barrier, $C_t>0$ is a numerical parameter and $\calT=\{(\alpha,\beta,\gamma) :  \alpha,\beta,\gamma \in \calD, \alpha \neq \beta \neq \gamma\}$, and we consider $n=1$ and $n=2$. 

As we shall see below, $\psi_s^{(1)}$ has the advantage that the effective potential for charged walls has minima that remain at the spontaneous polarization states but the drawback of presenting discontinuities in its derivatives. On the other hand, with $\psi_s^{(2)}$ discontinuities in the derivatives are absent, but the zeros and minima of the effective potential for charged walls are shifted from the spontaneous polarization states.

The contribution $\tau(\bfvarphi)$ to $\psi_s^{(n)}(\bfvarphi)$ serves to avoid the spurious appearance of a third phase in a domain wall by penalizing the simultaneous occurrence of more than two phases.\footnote{Note that if one chooses to write the triple term as a product of the squares of the $\varphi_\alpha$ instead of their absolute values, then this term does not fulfill its function, which is to strictly keep, within domain walls, any third phase at zero.}   Despite the apparently complicated expression of $\psi_s^{(n)}$, its section $\psi_{180}(\varphi)=\psi_s^{(n)}\big((\varphi,1-\varphi,0,0,0,0)\big)$ associated with a 180\deg domain wall reduces to
\begin{equation} \label{eq:hpsi180}
\psi_{180}(\varphi)=4^n \snorm{\varphi}^n \snorm{1-\varphi}^n,
\end{equation}
which is a double-obstacle ($n=1$) or double-well ($n=2$) potential of unitary barrier. Similarly, the section $\psi_{90}(\varphi)=\psi_s^{(n)}\big((\varphi,0,1-\varphi,0,0,0)\big)$ associated with a 90\deg domain wall reads
\begin{equation} \label{eq:hpsi90}
\psi_{90}(\varphi)=4^n h_{90} \snorm{\varphi}^n \snorm{1-\varphi}^n.
\end{equation}

\paragraph{Governing equations}
 
Whereas Gauss' law \eqref{eq:gauss}$_1$ and Faraday's equation \eqref{eq:faraday}$_1$ apply without change, the main difference between our general kinetics model and the classical phase-field model resides in the evolution equation for the multi-phase field,%
\footnote{We note that, by construction of \eqref{eq:evol}, $\varphi_\alpha$ only evolves in the diffuse-interface region, where its gradient is non-zero but remains constant in the bulk phases. As we will report in a forthcoming publication, the numerical implementation of \eqref{eq:evol} furnishes a seamless transition between these two domains (i.e., $\nabla \varphi_\alpha=\mathbf{0}$ and $\nabla \varphi_\alpha\neq \mathbf{0}$).} which we define as
\begin{equation} \label{eq:evol}
\dot \varphi_\alpha=\sum_{\beta=1}^6  \sqrt{\snorm{\nabla \varphi_\alpha \cdot \nabla \varphi_\beta}}\, G_{\beta \alpha}(h_{\beta \to \alpha}) \quad \mbox{for} \quad \alpha=1 \ldots 6,
\end{equation}
where $G_{\alpha \beta}=G_{\beta\alpha} : \dsR \rightarrow \dsR$ is the kinetic function associated with the transformation between phases $\alpha$ and $\beta$, and $h_{\beta \to \alpha}$ denotes the relative variational derivative of ${\Psi}$,
\begin{equation} \label{eq:hba}
h_{\beta \to \alpha}\left(\bfe,\bfvarphi,\nabla^2\bfvarphi\right)=-\frac{\delta \Psi}{\delta \varphi_\alpha }+\frac{\delta \Psi}{\delta \varphi_\beta }=-\partderiv{\Psi}{\varphi_\alpha}+\partderiv{\Psi}{\varphi_\beta}+\divv\left( \partderiv{\Psi}{\nabla\varphi_\alpha}-\partderiv{\Psi}{\nabla \varphi_\beta}\right).
\end{equation}
Inserting \eqref{eq:psihat} in \eqref{eq:hba} yields
\begin{equation} \label{eq:hba2}
h_{\beta \to \alpha}\left(\bfe,\bfvarphi,\nabla^2\bfvarphi\right)= \bfe \cdot (\bfp_{\alpha}-\bfp_{\beta})- C_s \bigg( \partderiv{\psi_s}{\varphi_\alpha} - \partderiv{\psi_s}{\varphi_\beta} \bigg) + 2 p_0^2 C_g \big( \nabla^2 \varphi_\alpha-\nabla^2 \varphi_\beta \big).
\end{equation}
For tetragonal ferroelectrics, we distinguish between two types of transformations: the motion of 180\deg  and 90\deg domain walls. Hence, we take the function $G_{\alpha \beta}$ as $G_{180}$ for $\{\alpha,\beta\} \in \calI_{180}$ and $G_{90}$ for $\{\alpha,\beta\} \in \calI_{90}$. By the second law of thermodynamics, $G_{180}$ and $G_{90}$ must be odd functions that satisfy the constraints
\begin{equation} \label{eq:grestrict}
G_{180}(h) h \geq 0 \quad \mbox{and} \quad G_{90}(h) h \geq 0 \quad \mbox{for all}\; h \in \dsR.
\end{equation}
Indeed, as we will show in Section~\ref{sec:1dsolgkm}, these two functions specify the kinetic relations of 180\deg  and 90\deg domain walls in the sense of \eqref{eq:kr1} in the sharp-interface model. Condition \eqref{eq:grestrict}, like \eqref{eq:frestrict}, ensures the positivity of the dissipation in the phase-field model.

Equation \eqref{eq:evol} is used with an initial condition $\bfvarphi(\bfx,0)$ that satisfies\footnote{%
One typically uses an initial condition describing a given arrangements of domains, within each of which, one $\varphi_\alpha$ is set to one and the other $\{\varphi_\beta\}_{\beta \neq \alpha}$ are set to zero with sharp transitions between domains.%
}%
\begin{equation} \label{eq:initialphi}
\sum_{\alpha=1}^6 \varphi_\alpha(\bfx,0)=1 \quad \mbox{for all} \quad \bfx \in \Omega.
\end{equation}
Using that $G_{180}$ and $G_{90}$ are odd functions, one can show that \eqref{eq:evol} guarantees $\sum_{\alpha=1}^6 \dot{\varphi}_\alpha =0$, whereby \eqref{eq:initialphi} holds for all times if it holds for the initial condition.

\paragraph{Comparison with other phase-field models with nonlinear kinetics}

The evolution equation \eqref{eq:evol} is an extension to multiple phases of the \emph{hybrid model} introduced by \citet{Alber2013} in the setting of stress-driven phase transformations between two phases. In particular, in a domain wall between two phases, say $\alpha=1$ and $\beta=2$, for which $\varphi_2(\bfx,0)=1-\varphi_1(\bfx,0)$ and  $\varphi_{\gamma}(\bfx,0)=0$ for $\gamma=3 \ldots 6$, \eqref{eq:evol} reduces to%
\begin{empheq}[left=\empheqlbrace]{align} \label{eq:evol1phase}
\begin{aligned}
\dot \varphi_1& = \snorm{\nabla \varphi_1} G_{180}\left(h_{2 \to 1} \right), \quad \varphi_2=1-\varphi_1,\\
\varphi_\gamma&=0 \quad \mbox{for} \quad \gamma=3 \ldots 6.
\end{aligned}
\end{empheq}
\eqref{eq:evol1phase}$_1$ is akin to Equation~(1.3) of \citet{Alber2013} (there formulated in the context of stress-driven transformations). Note that the double-well potential proposed by \citet{Alber2013} is a variation of \eqref{eq:hpsi180}.

Similarly, \eqref{eq:evol1phase}$_1$ is reminiscent of the evolution equation (2.7) of \citet{Agrawal2015}, which derives from a conservation law for the number of interfaces. However, in their work, the free-energy density was regularized differently from ours and that of \citet{Alber2013}, without resort to a double-well potential but through a regularized Heaviside function. Further, \citet{Agrawal2015} included a source term in the evolution equation for $\varphi$ to account for the nucleation of new domains.  In our case, in the absence of such a source term, \eqref{eq:evol} only propagates domain walls, while for nucleation new domains shall be introduced explicitly according to an appropriate criterion, which we do not discuss here.

\subsection{Analytical solutions for 180\deg  and 90\deg domain walls \label{sec:1dsolgkm}}

In this section, we compute analytical traveling wave solutions for the propagation of straight 180\deg  and 90\deg domain walls as obtained from the general kinetics model. In doing so, we follow in one dimension a procedure outlined, e.g., in \citet{Fried1994} for Allen--Cahn-type models. In particular, we demonstrate the nonlinear kinetics of domain walls that the general kinetics model permits, and we compare our findings to the Allen--Cahn model.

\subsubsection{Neutral 180\deg domain wall profile and kinetics} \label{sec:180ngkm}

\begin{figure}
\centering
\includegraphics[width=1\textwidth]{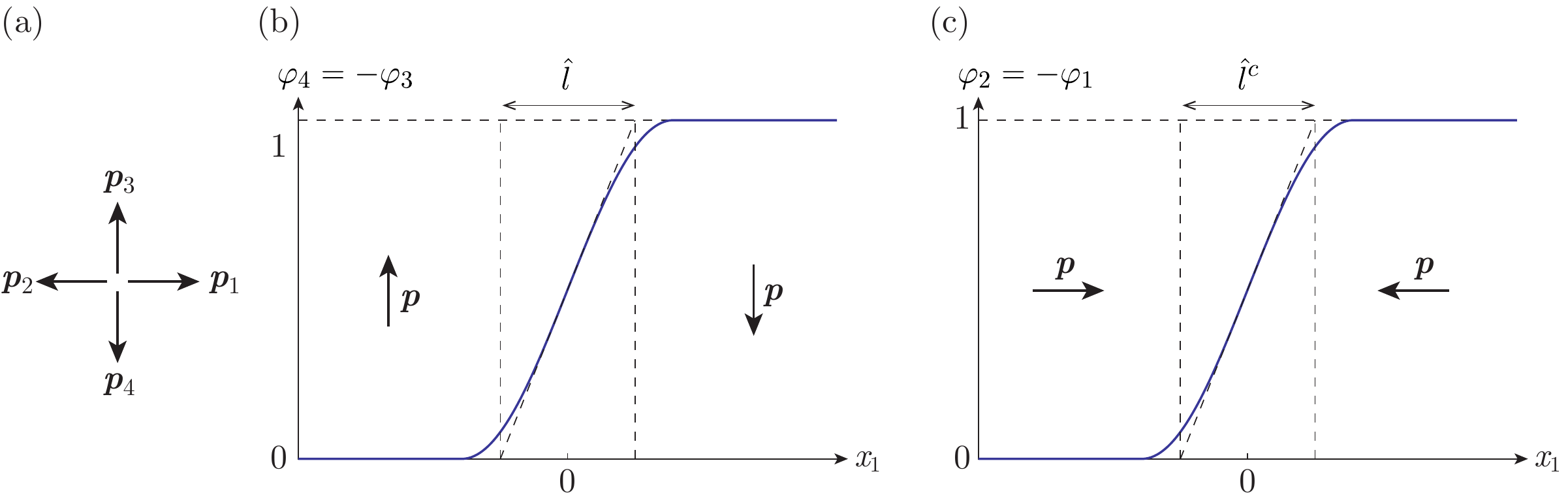}
\caption{(a) Basis of spontaneous polarization states. Phase-field profiles of two 180\deg domain walls: (b) charge-neutral wall, (c) charged wall.  The profile in (b) corresponds to \eqref{eq:gkmpro180} with $\varphi=\varphi_4$ and that in (c) to the solution of  \eqref{eq:gkmv180c} with $\varphi=\varphi_2$.}
\label{fig:gkmprofile180}
\end{figure}

We consider a straight 180\deg domain wall with polarization $\bfp(x_1,t)=p_2(x_1,t) \et_2$, separating the spontaneous polarizations $\bfp_3=p_0 \et_2$ at $x_1 \rightarrow - \infty$ and $\bfp_4= -p_0 \et_2$ at  $x_1 \rightarrow + \infty$ (see Figure~\ref{fig:gkmprofile180}(b)).
The phase fields not related to the two phases under consideration, i.e., $\varphi_\gamma$ for $\gamma\in \{1,2,5,6\}$ are initially assumed identically zero and, in view of \eqref{eq:evol}, remain zero for all time.%
\footnote{Note that, unlike with the classical Allen--Cahn evolution equation, \eqref{eq:evol} never leads to the nucleation of a new phase (say, phase $\alpha$) irrespective of the value of the electric field (and associated driving term $h_{\beta\rightarrow\alpha}$). This is due to the term $\nabla \varphi_\alpha$, which vanishes identically when the initial condition for $\varphi_\alpha$ is set to zero in the entire simulation domain. Accordingly, with this model a separate explicit nucleation criterion is required to account for nucleation events.}
Hence, only the phase fields $\varphi_3$ and $\varphi_4$, satisfying  $\varphi_3+\varphi_4=1$, are non-trivial. To simplify notation, we define $\varphi=\varphi_4$ to write the polarization profile as 
\begin{equation} \label{eq:pstar}
\bfp_{180}(\varphi)=p_0 (1 -2 \varphi) \et_2.
\end{equation}
As in Section~\ref{sec:180dw}, we can assume a uniform electric field $\bfe=e \et_2$ while satisfying Gauss' law \eqref{eq:gauss}$_1$.
The evolution equations \eqref{eq:evol} for $\varphi_3$ and $\varphi_4$ are identical and are written as an equation for $\varphi$:
\begin{equation} \label{eq:evolphi}
\dot{\varphi}=\snorm{\varphi_{,x_1}} G_{180}\big( h_{3\rightarrow4}(\varphi,\varphi_{,x_1x_1})\big),
\end{equation}
where the driving force \eqref{eq:hba2} reads
\begin{equation} \label{eq:h34}
h_{3\rightarrow4}(\varphi,\varphi_{,x_1x_1})=-2p_0 e-C_s  \psi'_{180}(\varphi)+4 C_g p_0^2 \varphi_{,x_1x_1}.
\end{equation}
As in Section~\ref{sec:1dsol}, we seek a traveling wave solution of \eqref{eq:evolphi} of the form
\begin{equation} \label{eq:phihat}
\varphi(x_1,t)=\cphi(x_1-v_{180} t)
\end{equation}
and introduce $\xi=x_1-v_{180} t$. Inserting \eqref{eq:phihat} into \eqref{eq:evolphi} yields
\begin{equation} \label{eq:v1}
-v_{180} \cphi_{,\xi} = \snorm{\cphi_{,\xi}} G_{180} \big( h_{3 \rightarrow 4}(\cphi,\cphi_{,\xi\xi}) \big).
\end{equation}
Without loss of generality, assuming that the solution satisfies $\cphi_{,\xi}\geq0$ for all $\xi$, \eqref{eq:v1} simplifies to
\begin{empheq}[left=\empheqlbrace]{align} \label{eq:v2}
\begin{aligned}
&v_{180}= G_{180} \bigg( 2p_0  e+{C_s} \psi'_{180}	(\cphi)-4 C_g p_0^2 \cphi_{,\xi \xi} \bigg) ,\\
&\cphi(-\infty)=0, \quad \cphi(+\infty)=1,\quad \cphi_{,\xi}(\pm \infty)=0,
\end{aligned}
\end{empheq}
where we have appended in \eqref{eq:v2}$_{2,3,4}$ the far-field boundary conditions on $\cphi$ consistent with the polarization states assumed at infinity. Note that \eqref{eq:v2}$_{2,3,4}$ are in fact two boundary conditions, since the vanishing spatial derivative of $\cphi$ at infinity is inherited from the former two conditions \eqref{eq:v2}$_{2,3}$. For a given applied electric field $e$, \eqref{eq:v2} can be solved for $\big(v_{180},\cphi(\xi) \big)$. We assume here that $G_{180}$ is monotonous on $\dsR$ and hence invertible, and we write its inverse as $G_{180}^{-1}$. Hence, \eqref{eq:v2}$_1$ is rewritten as 
\begin{equation} \label{eq:v3}
2 p_0 e+{C_s} \psi'_{180}(\cphi)-4 C_g p_0^2 \cphi_{,\xi \xi }  = G_{180}^{-1}(v_{180}).
\end{equation}
We compute the first integral of \eqref{eq:v3} by multiplying by $\cphi_{,\xi}$ and integrating with the condition at $-\infty$, which yields:
\begin{equation} \label{eq:v4}
\big[2p_0e - G_{180}^{-1}(v_{180}) \big] \cphi+C_s \psi_{180}(\cphi) -2 C_g p_0^2 (\cphi_{,\xi})^2=0,
\end{equation}
where we have used the fact that $\psi_{180}(0)=0$.
Further prescribing condition \eqref{eq:v2}$_{3,4}$ at $+\infty$ and noting that  $\psi_{180}(1)=0$, requires that 
\begin{equation} \label{eq:v5}
2p_0e=G_{180}^{-1}(v_{180}) \quad \Rightarrow \quad  v_{180}=G_{180}(2p_0e) ,
\end{equation}
and \eqref{eq:v4} reduces to an equation for $\cphi(\xi)$ only:
\begin{equation} \label{eq:phi}
C_s \psi_{180}(\cphi) -2 C_g p_0^2 (\cphi_{,\xi})^2=0.
\end{equation}
Equation \eqref{eq:phi} is similar to \eqref{eq:equipart} in the Allen--Cahn model. However, note that \eqref{eq:phi} holds for all $e$. Therefore, the profile is the same at equilibrium ($e=0$) and with an applied external electric field ($e\neq 0$). In particular, the polarization inside domains remains exactly equal to the spontaneous polarizations even under non-zero electric fields. This is in contrast to the Allen--Cahn model, whose equilibrium polarization profile is given by  \eqref{eq:equipart}. That equation, however, only provides an approximation of the profile in the presence of an electric field, which is valid only under condition \eqref{eq:esmall}.

Adopting definitions for the interfacial energy,
\begin{equation}
\Gamma=\int_{-\infty}^{+\infty} \Psi(\bfe,\check \bfvarphi,\nabla \check \bfvarphi) \dd \xi-\int_{-\infty}^{+\infty} W(\bfe,\bfp^\text{sharp}) \dd \xi,
\end{equation}
and the interfacial width,
\begin{equation}
l=1/\cphi_{,\xi}(0),
\end{equation}
equivalent to those of Section~\ref{sec:1dsol}, one can show using \eqref{eq:phi} that these quantities are given by expressions analogous to \eqref{eq:gamma3} and \eqref{eq:l2}, namely,
\begin{equation}
\Gamma= \eta p_0  \sqrt{2 C_s C_g}  \quad \mbox{and} \quad l=p_0 \sqrt{2 C_g/C_s},
\end{equation}
with $\eta=2 \int_0^1 \sqrt{\psi_{180}(\varphi)} \dd \varphi$. In particular, with $\psi_{180}$ given by \eqref{eq:hpsi180} we obtain $\eta=\pi/2$ for $n=1$ and $\eta=4/3$ for $n=2$.


By setting the origin to $\cphi(0)=1/2$, \eqref{eq:phi} forms an initial value problem for $\cphi(\xi)$, starting at $\xi=0$ and to be solved for both increasing and decreasing $\xi$.  Its integration yields
\begin{equation} \label{eq:gkmpro180}
\cphi(\xi)=\begin{cases} 
0 & \quad \mbox{for} \; \xi<-{\pi l}/{4},\\
\cos^2 \left(\displaystyle \frac{\pi}{4} - \frac{\xi}{l} \right) & \quad \mbox{for} \; -{\pi l}/{4}  \leq \xi \leq {\pi l}/{4}, \\
1 & \quad \mbox{for} \; \xi>{\pi l}/{4},
\end{cases}
\end{equation}
when using $\psi_s^{(1)}$ and
\begin{equation} \label{eq:gkmpro180n2}
\cphi(\xi)=\frac{\mathrm{e}^{4 \xi/l}}{1+\mathrm{e}^{4 \xi/l}},
\end{equation}
for $\psi_s^{(2)}$.
Note that with $\psi_s^{(1)}$ the diffuse interface representing the domain wall is localized in the interval $(-\pi l/4,\pi l/4)$.

With regards to kinetics, in \eqref{eq:v5}$_2$ $2 p_0 e=f$ corresponds to the driving traction \eqref{eq:df2} of the straight, neutral 180\deg domain wall in the sharp-interface model. Hence, $G_{180}$ directly provides the kinetic relation for 180\deg domain walls (i.e., \eqref{eq:kr1}  in the sharp-interface model). Therefore, the general kinetic model allows us to choose the kinetic relation arbitrarily (in the set of monotonously increasing odd functions that satisfy \eqref{eq:grestrict}). In particular, for modeling ferroelectrics, a physics-based choice for $G_{180}$ is to adopt the functional form \eqref{eq:kr} derived from experiments. In addition, note that for the neutral 180\deg domain wall, the kinetic relation is exactly given by \eqref{eq:v5}$_2$ for all $e$, whereas with the Allen--Cahn model the linear kinetic relation \eqref{eq:vel2} is only an approximation valid under the condition \eqref{eq:esmall}.
 

\subsubsection{Charged 180\deg domain wall profile and kinetics} \label{sec:180cgkm}

\begin{figure}
\centering
\includegraphics[width=1\textwidth]{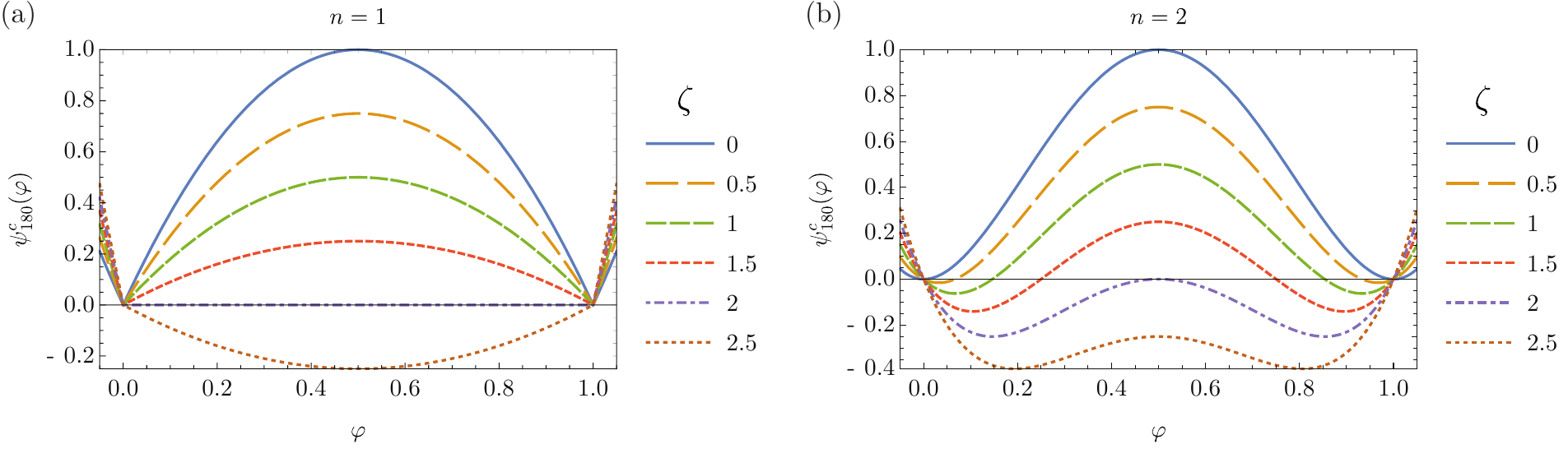}
\caption{Potential $\psi_{180}^c(\varphi)$ defined in \eqref{eq:psic180} for different values of $\zeta$ with (a) $n=1$ and (b) $n=2$ in the original double-well \eqref{eq:hpsi180}.}
\label{fig:hpsic}
\end{figure}

We consider the same charged domain wall as in Section~\ref{sec:180dwCharged}, i.e.,  $\bfp(x_1,t)=p_1(x_1,t) \et_1$, varying between $\bfp_1=p_0 \et_1$ at $x_1 \rightarrow - \infty$ and $\bfp_2= -p_0 \et_1$ at  $x_1 \rightarrow + \infty$ (see Figure~\ref{fig:gkmprofile180}(c)). For the same reasons as those invoked in Section~\ref{sec:180ngkm}, only $\varphi_1$ and $\varphi_2$ are non-identically zero, and they satisfy $\varphi_1+\varphi_2=1$. Using the single phase field $\varphi=\varphi_2$, we rewrite the polarization as
\begin{equation} \label{eq:pstar2}
\bfp_{180}^c(\varphi)=p_0(1-2 \varphi) \et_1.
\end{equation}
Further, with the same assumptions on the electric field and electric displacement as in Section~\ref{sec:180dwCharged}, the electric field is given by \eqref{eq:e180c}. We write the evolution equations for $\varphi_1$ and $\varphi_2$ as an equation for $\varphi$,
\begin{equation} \label{eq:evolphi2}
\dot{\varphi}=\snorm{\varphi_{,x_1}} G_{180}\big( h_{1\rightarrow2}(\varphi,\varphi_{,x_1x_1})\big).
\end{equation}
By combining \eqref{eq:hba2}, \eqref{eq:pstar2} and \eqref{eq:e180c}, $h_{1\rightarrow2}$ becomes
\begin{equation} \label{eq:h12}
h_{1\rightarrow2}(\varphi,\varphi_{,x_1x_1})=  -\frac{2 p_0 d}{\epsilon}-{C_s}
\underbrace{\left( \psi'_{180}(\varphi)-\frac{2p_0^2}{\epsilon C_s }\big(1-2 \varphi\big)\right)}_{{{\psi}^{c\,\prime}_{180}}(\varphi)}\,+\,4 C_g p_0^2 \varphi_{,x_1x_1},
\end{equation}
with $\psi_{180}^c(\varphi)$ the effective double-well potential for charged 180\deg domain walls defined by
\begin{equation} \label{eq:psic180}
\psi_{180}^c(\varphi)=\psi_{180}(\varphi)-2 \zeta \varphi (1-\varphi),
\end{equation}
with 
\begin{equation} \label{eq:zeta}
\zeta=\frac{p_0^2}{\epsilon C_s}=\frac{\eta l p_0^2}{\epsilon \Gamma}.
\end{equation}
Parameter $\zeta$ is equivalent to $\hzeta$ involved in \eqref{eq:fractionSmall} in the Allen--Cahn model and represents the ratio of the electric field $p_0/\epsilon$ induced by Gauss' law in a charged 180\deg domain wall to the numerical characteristic electric field $ e_c= \Gamma/(\eta l p_0)$. As is apparent from Figure~\ref{fig:hpsic}, the effective potential for $n=1$ is indeed a double-obstacle potential as long as $\zeta<2$ with minima and zeros coinciding at $\varphi=0$ and $\varphi=1$. In contrast, for $n=2$, minima of the potential are shifted inwards and we denote by $\varphi_m(\zeta)$ the smallest minimizer of $\psi_{180}^c$.

We look for a traveling wave solution of \eqref{eq:evolphi2} of the form 
\begin{equation} \label{eq:phihatc}
\varphi(x_1,t)=\cphi(x_1-v_{180}^c t)
\end{equation}
while assuming, without loss of generality, $\cphi_{,\xi} \geq 0$. Inserting \eqref{eq:phihatc} into \eqref{eq:evolphi2}, we obtain the equation for $(v_{180}^c,\cphi(\xi))$:
\begin{equation}\label{eq:gkmv180c}
    v_{180}^c= G_{180} \bigg(  \frac{ 2p_0 d}{\epsilon}+{C_s} \psi^{c\,\prime}_{180}	(\cphi)-4 C_g p_0^2 \cphi_{,\xi \xi}  \bigg).
\end{equation}

\paragraph{Case $n=1$}

In the case $n=1$, one can construct a solution that satisfies the far-field boundary conditions
\begin{equation} \label{eq:bcn1}
\cphi(-\infty)=0, \quad \cphi(+\infty)=1,\quad \cphi_{,\xi}(\pm \infty)=0.
\end{equation}

Following a derivation strictly analogous to that for neutral 180\deg domain walls, one can show that the solution of \eqref{eq:gkmv180c} supplemented by \eqref{eq:bcn1}  is a topological soliton moving at velocity $v_{180}^c=G_{180}(2p_0d/\epsilon)$ (with $2p_0d/\epsilon$ the sharp-interface driving traction \eqref{eq:df2} of a 180\deg charged wall). The polarization profile is given by \eqref{eq:gkmpro180}, if the width
\begin{equation} 
l^c=l \left(1-\frac{\zeta}{2}\right)^{-1/2}
\end{equation}
is substituted for $l$ and the interfacial energy is replaced by
\begin{equation} \label{eq:hgc}
\Gamma^c=\Gamma \left(1-\frac{\zeta}{2}\right)^{1/2}.
\end{equation}

While we see from \eqref{eq:hgc} that the regularization introduces a modification of the  interfacial energy between neutral and charged domain walls, this turns out not to be an issue in practice. From the point of view of the sharp-interface model, an accurate interfacial energy is important, as it contributes to the driving force $f$ through the effect of curvature in \eqref{eq:df2}. In practice we have $\zeta<2$, as required to have a double-obstacle potential. If $\zeta$ is low (typically $\zeta<0.2$) the change in interfacial energy remains less than 10\%. Therefore we expect an accurate account of the curvature contribution in \eqref{eq:df2}. By contrast, when $\zeta$ is large (typically $\zeta=1-1.5$), the artificial change in interfacial energy becomes larger (up to 50\% relative change for $\zeta=1.5$). However, when $\zeta$ is large, $\Gamma$ is small (see \eqref{eq:zeta}), and the contribution $\gamma \kappa$ to the driving force \eqref{eq:df2} becomes negligible ---compared to that of the bulk energies---almost everywhere except in local zones of high curvature. Hence, in this case the regularization introduces significant variations in $\Gamma$, but these are of little importance. In practice, the contribution of curvature of domain walls to the evolution of domains is mostly negligible in ferroelectric switching as well as in other solid-solid phase transformations. Therefore, the fact that the general kinetic model with $n=1$ allows us to take large values of $\zeta$ up to 1.5 is a significant advantage over the Allen--Cahn model and over general kinetic model with $n=2$ (discussed below), which both require $\zeta \ll 1$ (typically $\zeta<0.2$).

\paragraph{Case $n=2$}
For $n=2$ the far-field boundary conditions \eqref{eq:bcn1} do not permit to build a traveling wave solution. Instead, one needs to impose
\begin{equation} \label{eq:bcn2}
\cphi(-\infty)=\varphi_l, \quad \cphi(+\infty)=1-\varphi_l,\quad \cphi_{,\xi}(\pm \infty)=0
\end{equation}
with $\varphi_l$ a value that satisfies $\psi_{180}^{c \prime}(\varphi_l)>0$, i.e., $\varphi_m<\varphi_l<1/2$. Numerical simulations confirm that away from the interface, if $\varphi$ takes initial values at 0 and 1, it evolves toward the solution of (take $h_{1\rightarrow2}=0$ with $\varphi_{,x_1 x_1}=0$ in \eqref{eq:h12})
\begin{equation}
    \psi'_{180}(\varphi)=-2 p_0 e / C_s,
\end{equation}
where $e = \calO(p_0/\epsilon)$ is the locally uniform (away from the interface) electric field. This implies that $\zeta$ shall satisfy $\zeta \ll 1$ to ensure polarization remains close to the spontaneous polarization states. The traveling wave solution that can be constructed with $\varphi$ varying from $\varphi_l$ to $1-\varphi_l$, although not connecting $\varphi=0$ to $\varphi=1$,  has exactly the velocity expected from the target sharp interface model, $v_{180}^c= G_{180}(2 p_0 d / \epsilon)$.  In numerical simulations, diffuse interfaces evolve with the velocity predicted here and connect to the values of $\varphi$ within phases through transient profiles.

\subsubsection{90\deg domain walls: profile and kinetics\label{sec:gkm90}}

\paragraph{Neutral domain wall}

\begin{figure}
\centering
\includegraphics[width=1\textwidth]{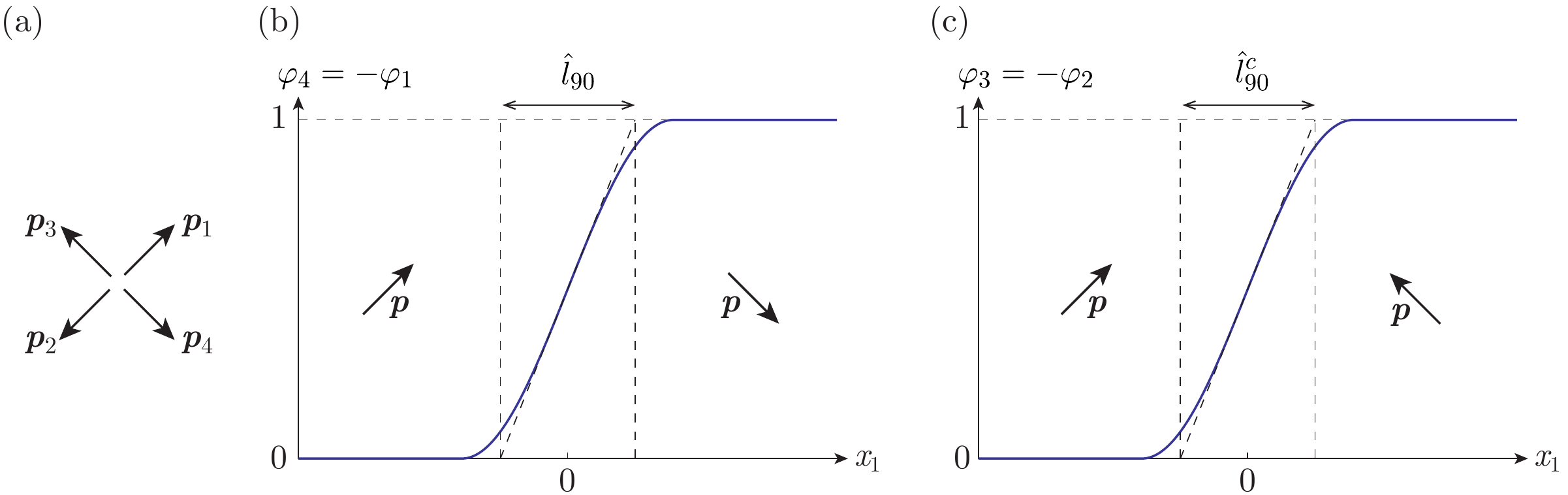}
\caption{(a) Basis of spontaneous polarization states. Phase-field profiles of two 90\deg domain walls: (b) charge-neutral wall; (c) charged wall, with properties discussed in Section~\ref{sec:gkm90}.}
\label{fig:gkmprofile90}
\end{figure}

As in Section~\ref{sec:90dw}, the basis of spontaneous polarizations $\{ \bfp_\alpha \}_{\alpha \in (1,6) }$ defined in \eqref{eq:pbasis} is rotated by $+\pi/4$ around the $\et_3$-axis. Consider, e.g., the same 1D head-to-tail domain wall as in Section~\ref{sec:90dw}, i.e., between $\bfp_1=p_0/\sqrt{2}(\et_1+\et_2)$ at $x_1 \rightarrow - \infty$ and $\bfp_4=p_0/\sqrt{2}(\et_1-\et_2)$ at $x_1 \rightarrow + \infty$ (see Figure~\ref{fig:gkmprofile90}(b)).
This wall involves $\varphi_1$ and $\varphi_4$ satisfying $\varphi_1+\varphi_4=0$, while the other phase fields are identically zero. For simplicity we define $\varphi=\varphi_4$. A specialty of the polarization profile obtained from the general kinetics model is that the rotation of the polarization vector across the wall is predetermined by the decomposition \eqref{eq:phat}, i.e., 
\begin{equation}
\overset{\ast}{\bfp}_{90}(\varphi)=\frac{p_0}{\sqrt{2}} \big[ \et_1 + (1 - 2 \varphi) \et_2 \big],
\end{equation}
whereby $\bfp$ rotates with a constant component along $\et_1$. This is in contrast to the rotation of the polarization vector resulting from the Allen--Cahn model, which depends on the multi-well potential $\opsi_s$ given by \eqref{eq:opsis}, in particular on the parameter $\hchi$ introduced in \eqref{eq:hchi}, which determines the location of the 90\deg barrier.

As a consequence, whereas in Section~\ref{sec:90dw}  the computed properties (wall width and interfacial energy) of the neutral 90\deg domain wall are only approximations relying on the assumption that the $p_1$-component remains constant (a reasonable assumption for $\hchi=0.5$), for the general kinetics model these same properties characterize the neutral 90\deg domain wall exactly.

Under the application of a uniform electric field $\bfe=e \et_2$ and with a derivation in all points analogous to that developed in Section~\ref{sec:180ngkm}, we conclude that the 90\deg domain wall can be described by a traveling wave solution,
\begin{equation}
\varphi(x_1,t)=\cphi(x_1-v_{90} t),
\end{equation}
where $\xi=x_1-v_{90} t$ and $v_{90}$ is the velocity of the 90\deg domain wall. Under the assumption that function $G_{90}$ is monotonous on $\dsR$, one can show that, akin to \eqref{eq:v5}, 
\begin{equation}
v_{90}=G_{90}(\sqrt{2} p_0 e),
\end{equation}
where $\sqrt{2} p_0 e$ corresponds to the driving traction \eqref{eq:df2} associated with the 90\deg configuration under consideration. Again, the polarization profile is given by \eqref{eq:gkmpro180} and \eqref{eq:gkmpro180n2} with a width $l_{90}=l/\sqrt{h_{90}}$ replacing $l$ and an interfacial energy $\Gamma_{90}=\sqrt{h_{90}} \Gamma$, which can be set to its physical value through a proper choice of $h_{90}$.

\paragraph{Charged domain wall}

The transition from the neutral to the charged 90\deg domain wall (see Figure~\ref{fig:gkmprofile90}(c)) shows the same features as the analogous transition for the 180\deg domain wall. We discuss specifically the case $n=1$, for which the kinetic relation is exactly given by the function $G_{90}$, provided that $\zeta<2\sqrt{2}h_{90}$ is satisfied (like $\zeta<2$ for 180\deg walls). The modified values of interfacial width and energy due to the regularization with $n=1$ are 
\begin{equation}
l^c_{90}=l_{90} \left(1-\frac{\zeta}{2 \sqrt{2} h_{90}}\right)^{-1/2} \qquad \mbox{and} \qquad \Gamma^c_{90}=\Gamma_{90} \left(1-\frac{\zeta}{2 \sqrt{2} h_{90}}\right)^{1/2}.
\end{equation}
%

\subsection{Summary of the properties of 180\deg  and 90\deg domain walls\label{sec:gkmsum}}

\paragraph{Interface energy and width}

We have computed exactly the interfacial energy and width of the different types of domain walls obtained with the general kinetics model. The expressions are analogous to those of the Allen--Cahn model. In particular, the choice of $C_s$ and $C_g$ allows us to set independently the interfacial energy and width (i.e., the regularization length) of 180\deg domain walls, while $h_{90}$ determines the ratio of interfacial energy of 90\deg  and 180\deg domain walls. Furthermore, the expressions for the interfacial properties of walls remain exact under an applied electric field, in contrast to the Allen--Cahn model for which one only has approximate expressions of interfacial properties when domain walls are not in equilibrium. This is notably related to the fact that, with the general kinetics model, values of the spontaneous polarization within domains are unchanged in the presence of an electric field. 

Lastly, a special feature of the general kinetics model is that for 90\deg domain walls the rotation of the polarization occurs with one of its components kept constant, as predetermined by the decomposition \eqref{eq:phat}. This is in contrast to what happens with the Allen--Cahn model, where the rotation of $\bfp$ depends on the location of the 90\deg barrier, given by $\hchi$ in the multi-well potential.

\paragraph{Kinetics}

We have found that for straight domain walls the general kinetics model allows us to prescribe independently the kinetics of 180\deg  and 90\deg domain walls and that these kinetic relations can be nonlinear. The kinetic relations \eqref{eq:kr1} between the velocity of the interface (velocity of the traveling wave solutions) and the driving traction (computed from the values of the fields away from the diffuse interface) are defined through the functions $G_{180}$ and $G_{90}$ for 180\deg  and 90\deg domain walls, respectively. 

Interestingly, in the general kinetics model, the kinetic relation of straight walls is exactly given by $G_{180}$ and $G_{90}$ irrespective of the regularization length $l$, provided that $\zeta < 2$ and  $\zeta < 2 \sqrt{2} h_{90}$ are satisfied (i.e, $l$ is smaller than a critical length determined by the material parameters). This is in contrast to the Allen--Cahn model, for which the accuracy of the linear kinetic relations \eqref{eq:vel2} and \eqref{eq:kin90} for straight 180\deg  and 90\deg domain walls decreases as the regularization length $l$ increases (because, as $l$ increases with constant material parameters, $e_c$ in \eqref{eq:esmall} decreases, which reduces the accuracy of \eqref{eq:vel2} and \eqref{eq:kin90}). 

As will be discussed in Part~II of this study, when it comes to \textit{curved domain walls}, the general kinetics model allows us to introduce the kinetic relations via $G_{180}$ and $G_{90}$ in an exact sense only in the limit of $l \rightarrow 0$ (see also the asymptotic analysis of \citet{Alber2013} for the phase-field model that inspired the present general kinetics model).

\section{Conclusion\label{sec:conc}}

We have formulated a phase-field model for ferroelectrics that permits domain wall evolution with nonlinear kinetics, which we term the \emph{general kinetics model}. Nonlinear domain wall kinetics is a crucial feature for properly accounting for rate effects in ferroelectric switching and is nonetheless absent from existing diffuse-interface ferroelectrics models. Seen in the general class of phase-field models for structural transformations (which includes switching between ferroelectric variants, twinning, solid-solid phase transformations), the general kinetics model is, to the best of our knowledge, the first diffuse-interface model with arbitrary nonlinear interface kinetics formulated for systems with multiple (i.e., more than two) phases. It has been devised as an extension to multiple phases of the \emph{hybrid model} of \citet{Alber2013} (adapted here to the setting of ferroelectric switching).

To shed light on the new features of the general kinetics model, as compared to the classical Allen--Cahn-based phase-field model, we have worked in the simplified setting of rigid ferroelectrics, an assumption which only accounts for the electrostatic contribution to domain evolution and neglects the mechanical one.\footnote{Note that, building upon existing phase-field models by \citet{Schrade2013,Schrade2014}, which account for electro-mechanical coupling, the extension of our model to deformable ferroelectrics is straightforward. Indeed, the passage from an Allen--Cahn-type phase-field model that includes electro-mechanical coupling (such as \citet{Schrade2013,Schrade2014})  to a general kinetic model is conceptually the same as what we have presented in Section~\ref{sec:gkm} in the setting of rigid ferroelectrics.} In this setting, we have compared two regularized models: the classical Allen--Cahn-based phase-field model and the newly introduced general kinetics model. By computing analytical traveling wave solutions for the propagation of straight 180\deg and 90\deg domain walls, both neutral and charged, we have established the following properties of the two phase-field models:
\begin{enumerate}
\item
\emph{Interfacial width and energy}: For both diffuse-interface models, the numerical parameters associated with the regularization of the electric enthalpy permit to set independently the interfacial energy of 180\deg  and 90\deg domain walls and the width of 180\deg domain walls. While the former are material parameters characterizing the walls, we see the latter as a \emph{numerical regularization length}, whose choice is limited by practical constraints. With the Allen--Cahn model, interfacial width and energy weakly depend on the applied electric field and remain approximately constant in the regime \eqref{eq:esmall} of $| \bfe | \ll \he_c$, with $\he_c=\hG/(\heta \hatl p_0)$ a numerical characteristic electric field dependent on the regularization length $\hatl$. By contrast, with the general kinetics model, the interfacial width and energy of neutral domain walls are constant and independent of the applied electric field. As for charged domain walls (which encounter nonuniform electric fields induced by Gauss' law), both models require, for wall properties to remain approximately independent of the regularization length, that the magnitude of the electric field associated with Gauss' law remains small compared to a characteristic electric field. This translates into an upper bound on $\hatl$ and $l$, which reads as ${\heta \hatl p_0^2}/({\epsilon \hG}) \ll 1$ (see \eqref{eq:fractionSmall}) for the Allen--Cahn model and ${\eta l p_0^2}/({\epsilon \Gamma}) \ll 1$ for the general kinetics model (with quantitative estimates given in Sections~\ref{sec:180cgkm} and \ref{sec:gkm90}).

\item
\emph{Kinetics of domain walls}: For the motion of straight 180\deg  and 90\deg domain walls, we have shown that the Allen--Cahn formulation furnishes a regularization of a sharp-interface model with a linear kinetic relation between wall velocity and associated driving traction. In addition, while the kinetic coefficient associated with one type of domain walls (e.g., 180\deg domain wall) can be freely set with a proper choice of the inverse mobility $\mu$ in \eqref{eq:ac}, that of the other type of wall (e.g., 90\deg domain wall) is automatically prescribed by the ratio of interfacial energies of the two types of domain walls. Furthermore, the Allen--Cahn model only furnishes an \emph{approximately} linear kinetics behavior valid in the regime where both \eqref{eq:esmall} and \eqref{eq:fractionSmall} are satisfied\footnote{%
In practice, the magnitude $p_0/\epsilon$ of the electric field that results from Gauss' law in charged walls in significantly larger than that of externally applied electric fields, so that \eqref{eq:fractionSmall} is the more stringent condition}.%

The general kinetics model, by contrast, provides a regularization of the sharp-interface model with arbitrary kinetic relations given by the functions $G_{180}$ and $G_{90}$ for 180\deg  and 90\deg domain walls, respectively. More specifically, for straight walls of all types, we have shown with traveling wave solutions that the relations between wall velocities and associated driving tractions are exactly given by $G_{180}$ and $G_{90}$, provided that the (little restrictive) conditions $\zeta < 2$ and  $\zeta < 2 \sqrt{2} h_{90}$ are satisfied. Overall, they again furnish upper bounds on the regularization length $l$.
\end{enumerate}

In the present article, we have focused on analytically tractable features of the Allen--Cahn and general kinetics models. For that reason, we have restricted our attention to the properties and motion of straight domain walls. With this focus, we made the important clarification that with the Allen--Cahn model, the target kinetics is attained only in the limit of $\hatl \rightarrow 0$, while the general kinetics model furnishes exactly the chosen kinetics, irrespective of the regularization length, provided the latter satisfies specific upper bounds. We point out that, although our discussion was specific to ferroelectric domain wall motion, the same general kinetics model bears potential for other applications involving structural transformations.

In a follow-up publication, we will report a numerical implementation of the general kinetics model and address the propagation of curved walls as well as the behavior of triple and quadruple points where multiple phases meet.

\section*{Acknowledgment}

The support from the Swiss National Science Foundation (SNSF) is gratefully acknowledged. The authors thank L.~Hennecart, R.~Indergand, and V.~Kannan for stimulating discussions. We thank the anonymous reviewers for their useful comments on the manuscript.

 \newpage

\appendix

\section{Experimental evidence on domain wall dynamics \label{app:DWdynamics}}

\paragraph{180\deg domain walls} Experimental and theoretical works have probed the kinetics of domain walls both in bulk single-crystals and in thin films (for an extensive review of these works, see \citealt{Tagantsev2010}). Here, we summarize results pertaining only to bulk BaTiO$_3$. The velocity of non-ferroelastic 180\deg domains walls was measured as a function of the electric field in single-crystal BaTiO$_3$ by \citet{Miller1958,Miller1959,Miller1960,Savage1960} for electric fields in the range 0.1 to 2$\kVcm$ and by \citet{Stadler1963} at large fields between 2 and 450$\kVcm$. These experiments were performed on plate-like samples with polarization and electric field oriented along the out-of-plane $\et_z$-direction. As a result, the electric field $\bfe=e \et_z$ is uniform throughout the sample and the driving traction given by \eqref{eq:df1} reduces to $f=2  p_0e$ for an interface with negligible curvature. For electric fields in the range $0.1$ to $2\kVcm$, an inverse exponential dependence of the domain wall velocity on the electric field was inferred:
\begin{equation}
V_{180}(e)=V_l \exp(-e_a/e),
\end{equation}
where $V_l$ is a characteristic velocity of the order of $100\, \mathrm{cm/s}$, and the \emph{activation field} $e_a$ is typically about $4\, \mathrm{kV/cm}$ at room temperature \citep{Miller1960}. The exact value of these parameters depends on temperature, sample thickness, the presence of defects, and the nature of the electrodes.  These observations suggest that for 180\deg domain walls in the range $f \leq 1 \times 10^5 \,\mathrm{J/m^3}$ (cf. $p_0=0.26 \, \mathrm{C/m^2}$), the driving traction follows a kinetic relation of inverse-exponential type, given by
\begin{equation} \label{eq:kinLF}
 V_{180}(f)=\sgn(f) V_l \exp (-f_a/|f|),
\end{equation}
where $f_a=2 p_0 e_a$ is an \emph{activation driving traction}. Above 2$\kVcm$ the velocity was found to follow a power law $V_{180} \propto e^{\theta}$ with $\theta=1.4$ \citep{Stadler1963}, which indicates for $f > 1 \times 10^5 \,\mathrm{J/m^3}$ a kinetic relation of the form
\begin{equation}  \label{eq:kinHF}
 V_{180}(f)=\sgn(f) V_h \left(\frac{|f|}{f_a}\right)^{\theta},
\end{equation}
with $V_{h}$ a characteristic velocity for large driving tractions. \\

\paragraph{90\deg domain walls}  While experimental data for the motion of 90\deg domain walls in BaTiO$_3$ are lacking, observations of the kinetics of ferroelastic domain walls in other materials (Rochelle salt, galodinium molybdate, potassium dihydrogen phosphate as reviewed in \citet{Tagantsev2010}, Section 8.3.3) point to a different kinetic relation specific to ferroleastic domain walls. Indeed, irrespective of the material, it appears that ferroelastic walls remain static for applied electric fields smaller than a threshold field, above which their velocity evolves linearly with the field. This suggests a threshold-type linear relation:
\begin{equation}
 V_{90}(f)=\begin{cases}
0 \quad & \mbox{for} \quad  |f| < f_0, \\
\sgn(f) k (|f|-f_0)  &  \mbox{for} \quad |f| \geq f_0,
\end{cases}
\end{equation}
where $f_0$ is a threshold driving traction and $k$ a kinetic coefficient. At the same time, \emph{ab initio} calculations by \cite{Liu2016} have shown that 90\deg domain walls in PbTiO$_3$ exhibit the same two-regime kinetics as that observed for 180\deg domain walls in BaTiO$_3$ and described by \eqref{eq:kinLF} and \eqref{eq:kinHF} in the regimes of small and large electric fields, respectively. While we unfortunately do not have a full set of data for the kinetics of the different kinds of domain walls in one particular material, current observations clearly indicate that the kinetics of domain walls is nonlinear and dependent on the type of domain wall (i.e., non-ferroelastic vs. ferroelastic). 

\newpage

\bibliography{ferro_laurent}

\end{document}